%% file: methods.tex
\documentclass[twocolumn,prd,superscriptaddress,altaffilletter,nofootinbib]{revtex4-1}
\usepackage{amsmath}
\usepackage{amssymb}
\usepackage{amsfonts}
\usepackage{comment}
\usepackage{graphicx}
\usepackage{hyperref}
\usepackage[nolist]{acronym}
\usepackage[usenames,dvipsnames]{xcolor}
\usepackage{xspace}
\usepackage{tikz}
\usepackage[squaren]{SIunits}
\usetikzlibrary{arrows,shapes,trees,decorations.pathreplacing}
\allowdisplaybreaks

\newcommand{\bigdogfap}{\ensuremath{5.4 \times 10^{-9}}}
\newcommand{\bigdogsigma}{\ensuremath{5.7 \sigma}}
\newcommand{\bigdoghsnr}{14.7}
\newcommand{\bigdoglsnr}{9.4}
\newcommand{\bigdoghxisq}{1.4}
\newcommand{\bigdoglxisq}{1.5}
\newcommand{\bigdoglikelihood}{39.6}
\newcommand{\bigdogmchirp}{4.7}
\newcommand{\bigdogfar}{\ensuremath{1.1 \times 10^{-14}}}
\newcommand{\bigdogifar}{\ensuremath{2.7 \times 10^6}}

\newcommand{\likehood}{\mathcal{L}}
\newcommand{\snr}[1]{\rho_{\rm{#1}1 }} 
\newcommand{\chisq}[1]{\xi^2_{\rm{#1} 1}}
\newcommand{\detectors}{\{\rm{H}1,\rm{L}1\}}
\newcommand{\horizons}{\{D_{\rm{H}1},D_{\rm{L}1}\}}
\newcommand{\params}{\bar{\theta}}
\newcommand{\cprob}[2]{P ( #1 \mid #2 ) } 
\newcommand{\sh}{\mathrm{signal}} 
\newcommand{\nh}{\mathrm{noise}} 

\newcommand{\gstlal}{GstLAL\xspace}

\newcommand{\pipeline}{GstLAL-based inspiral pipeline}
\newcommand{\gracedb}{GraceDB}

\newcommand{\mchirp}{\ensuremath{\mathcal{M}}\xspace}
\newcommand{\msun}{\ensuremath{\mathrm{M}_{\odot}}\xspace}
\newcommand{\Mc}{\ensuremath{\mathcal{M}}}

\newcommand{\NB}{\ensuremath{N_B}}
\newcommand{\NT}{\ensuremath{N_T}}

\DeclareMathOperator{\order}{\mathcal{O}}

\begin{document}

\input{title.tex}

\date{\today}
\input{abstract.tex}
\maketitle

\section{Introduction}\label{s:intro}

	\input{intro.tex}

\section{Matched Filtering Input} \label{sec:initial}

	\input{sec2intro.tex}

	\subsection{Data Acquisition} \label{ss:data}

		\input{data.tex}

	\subsection{PSD} \label{ss:psd}

		\input{psd.tex}
	\subsection{Data Conditioning} \label{ss:data_cond}

		\input{datacond.tex}

	\subsection{Template Bank Decomposition} \label{ss:bank}

		\input{bank.tex}

\section{Event Identification Stage} \label{sec:ident}

		\input{eventid.tex}

		\subsection{Matched filtering and the LLOID method} \label{ss:matched}

		\input{matchedfiltering.tex}

		\subsection{Triggers} \label{ss:triggering}

		\input{triggering.tex}

		\subsection{Signal-based vetoes} \label{ss:sigveto}

		\input{autoveto.tex}

		\subsection{Coincidence} \label{ss:coinc}

		\input{coincidence.tex}

	\subsection{Event Ranking} \label{ss:rank}

		\input{likelihood.tex}

	\subsection{Event Clustering} \label{ss:cluster}

		\input{datareduction.tex}

\section{Event Processing and Sensitivity Estimation} \label{sec:final}

	\input{eventprocessing.tex}

	\subsection{Event Significance Estimation} \label{ss:sig}

		\input{significance.tex}

	\subsection{Generating Alerts} \label{ss:alerts}

		\input{alerts.tex}

	\subsection{Software injections} \label{ss:injections}

		\input{injections.tex}



\section{Conclusion} \label{sec:conclusion}

	\input{conclusion.tex}

\section{Acknowledgements} \label{sec:ack}

	\input{acknowledgements.tex}

\appendix \label{sec:appendix}

 \input{appendix1.tex}

\input{methods.bbl}

\input{acronyms.tex}
\end{document}

%% file: title.tex
\title{Analysis Framework for the Prompt Discovery of Compact Binary Mergers in Gravitational-wave Data}

\author{Cody Messick}
\email{Cody.Messick@ligo.org}
\affiliation{Department of Physics, The Pennsylvania State University, University Park, PA 16802, USA}
\affiliation{Institute for Gravitation and the Cosmos, The Pennsylvania State University, University Park, PA 16802, USA}

\author{Kent Blackburn}
\affiliation{LIGO Laboratory, California Institute of Technology, MS 100-36, Pasadena, California 91125, USA}

\author{Patrick Brady}
\affiliation{Leonard E.\ Parker Center for Gravitation, Cosmology, and Astrophysics, University of Wisconsin-Milwaukee, Milwaukee, WI 53201, USA}

\author{Patrick Brockill}
\affiliation{Leonard E.\ Parker Center for Gravitation, Cosmology, and Astrophysics, University of Wisconsin-Milwaukee, Milwaukee, WI 53201, USA}

\author{Kipp Cannon}
\affiliation{Canadian Institute for Theoretical Astrophysics, 60 St. George Street, University of Toronto, Toronto, Ontario, M5S 3H8, Canada}
\affiliation{RESCEU, University of Tokyo, Tokyo, 113-0033, Japan}

\author{Romain Cariou}
\affiliation{D\'{e}partement de physique, \'{E}cole Normale Sup\'{e}rieure de Cachan, Cachan, France}

\author{Sarah Caudill}
\affiliation{Leonard E.\ Parker Center for Gravitation, Cosmology, and Astrophysics, University of Wisconsin-Milwaukee, Milwaukee, WI 53201, USA}

\author{Sydney J. Chamberlin}
\affiliation{Department of Physics, The Pennsylvania State University, University Park, PA 16802, USA}
\affiliation{Institute for Gravitation and the Cosmos, The Pennsylvania State University, University Park, PA 16802, USA}

\author{Jolien D. E. Creighton}
\affiliation{Leonard E.\ Parker Center for Gravitation, Cosmology, and Astrophysics, University of Wisconsin-Milwaukee, Milwaukee, WI 53201, USA}

\author{Ryan Everett}
\affiliation{Department of Physics, The Pennsylvania State University, University Park, PA 16802, USA}
\affiliation{Institute for Gravitation and the Cosmos, The Pennsylvania State University, University Park, PA 16802, USA}

\author{Chad Hanna}
\affiliation{Department of Physics, The Pennsylvania State University, University Park, PA 16802, USA}
\affiliation{Department of Astronomy and Astrophysics, The Pennsylvania State University, University Park, PA 16802, USA}
\affiliation{Institute for Gravitation and the Cosmos, The Pennsylvania State University, University Park, PA 16802, USA}

\author{Drew Keppel}
\affiliation {Albert-Einstein-Institut, Max-Planck-Institut f\"ur Gravi\-ta\-tions\-physik, D-30167 Hannover, Germany }

\author{Ryan N.\ Lang}
\affiliation{Leonard E.\ Parker Center for Gravitation, Cosmology, and Astrophysics, University of Wisconsin-Milwaukee, Milwaukee, WI 53201, USA}

\author{Tjonnie G. F. Li}
\affiliation{Department of Physics, The Chinese University of Hong Kong, Shatin, New Territories, Hong Kong, China}

\author{Duncan Meacher}
\affiliation{Department of Physics, The Pennsylvania State University, University Park, PA 16802, USA}
\affiliation{Institute for Gravitation and the Cosmos, The Pennsylvania State University, University Park, PA 16802, USA}

\author{Alex Nielsen}
\affiliation {Albert-Einstein-Institut, Max-Planck-Institut f\"ur Gravi\-ta\-tions\-physik, D-30167 Hannover, Germany }

\author{Chris Pankow}
\affiliation{Center for Interdisciplinary Exploration and Research in Astrophysics (CIERA) and Department of Physics and Astronomy, Northwestern University, 2145 Sheridan Road, Evanston, IL 60208, USA}

\author{Stephen Privitera}
\affiliation{Albert-Einstein-Institut, Max-Planck-Institut f{\"u}r Gravitationsphysik, D-14476 Potsdam-Golm, Germany}

\author{Hong Qi}
\affiliation{Leonard E.\ Parker Center for Gravitation, Cosmology, and Astrophysics, University of Wisconsin-Milwaukee, Milwaukee, WI 53201, USA}

\author{Surabhi Sachdev}
\affiliation{LIGO Laboratory, California Institute of Technology, MS 100-36, Pasadena, California 91125, USA}

\author{Laleh Sadeghian}
\affiliation{Leonard E.\ Parker Center for Gravitation, Cosmology, and Astrophysics, University of Wisconsin-Milwaukee, Milwaukee, WI 53201, USA}

\author{Leo Singer}
\affiliation{NASA/Goddard Space Flight Center, Greenbelt, MD 20771, USA}

\author{E. Gareth Thomas}
\affiliation{University of Birmingham, Birmingham, B15 2TT, UK}

\author{Leslie Wade}
\affiliation{Department of Physics, Hayes Hall, Kenyon College, Gambier, Ohio 43022, USA}

\author{Madeline Wade}
\affiliation{Department of Physics, Hayes Hall, Kenyon College, Gambier, Ohio 43022, USA}

\author{Alan Weinstein}
\affiliation{LIGO Laboratory, California Institute of Technology, MS 100-36, Pasadena, California 91125, USA}

\author{Karsten Wiesner}
\affiliation {Albert-Einstein-Institut, Max-Planck-Institut f\"ur Gravi\-ta\-tions\-physik, D-30167 Hannover, Germany }

%% file: abstract.tex
\begin{abstract}
We describe a stream-based analysis pipeline to detect gravitational waves from
the merger of binary neutron stars, binary black holes, and
neutron-star--black-hole binaries within $\sim 1$ minute of the arrival of the
merger signal at Earth.  Such low-latency detection is crucial for the prompt
response by electromagnetic facilities in order to observe any fading
electromagnetic counterparts that might be produced by mergers involving at
least one neutron star.  Even for systems expected not to produce counterparts,
low-latency analysis of the data is useful for deciding when not to point
telescopes, and as feedback to observatory operations. Analysts using this
pipeline were the first to identify GW151226, the second gravitational-wave
event ever detected.  The pipeline also operates in an offline mode, in which
it incorporates more refined information about data quality and employs acausal
methods that are inapplicable to the online mode. The pipeline's offline mode
was used in the detection of the first two gravitational-wave events, GW150914
and GW151226, as well as the identification of a third candidate, LVT151012.
\end{abstract}

%% file: intro.tex
The field of gravitational-wave astronomy has come to life in a spectacular
way, with the first detections of gravitational waves on September 14,
2015~\cite{abbott2016observation} and December 26,
2015~\cite{abbott2016gw151226} by the two detectors of the Laser Interferometer
Gravitational-wave Observatory (LIGO)~\cite{aasi2015advanced}. These detectors
are currently undergoing further commissioning and will reach design
sensitivity in the next few years. Additionally, they will be joined by a
network of gravitational-wave observatories that include Advanced
Virgo~\cite{acernese2015advanced}, KAGRA~\cite{aso2013interferometer}, and a
third LIGO observatory in India~\cite{LigoIndia}. We expect this network to
bring more observations of binary black hole mergers~\cite{RATESPAPER}, as well
as binary neutron star (BNS) and neutron-star--black-hole (NSBH)
mergers~\cite{abadie2010predictions}.

As we enter the era of gravitational-wave astronomy, the need for low-latency
analyses becomes critical. Gravitational waves from BNS and NSBH mergers are
expected to be paired with electromagnetic emission and
neutrinos~\cite{singer2014first, abbott2016emfollow,
adrian2016neutrino}. Gravitational-wave-triggered electromagnetic observations
may lead to the detection of prompt short gamma-ray bursts and high-energy
neutrinos within seconds, followed by X-ray, optical, and radio afterglows days
to years later. Multimessenger observations will aid in our understanding of
astrophysical processes and increase our search
sensitivity~\cite{singer2014first, GWEMAlerts}. Additionally, even in the
absence of a counterpart, the rapid identification of gravitational waves has a
number of benefits. Low-latency detection allows us to provide feedback to
commissioners when search sensitivity drops unexpectedly, helping to return the
detector to its nominal state~\cite{abbott2016detchar}. Furthermore, upon
identification of a candidate, we can submit timely requests to minimize
detector changes in order to gather enough data to reliably estimate the search
background and perform followup calibration measurements.

In this work, we present the \gstlal{}-based inspiral pipeline, a
gravitational-wave search pipeline based on the \gstlal{}
library~\cite{gstlal}, and derived from GStreamer~\cite{gstreamer} and the LIGO
Algorithm Library~\cite{lalsuite}.  The pipeline can operate in a low-latency
mode to ascertain whether a gravitational-wave signal is present in data,
provide point estimates for the binary parameters, and estimate event
significance. Analysts running the low-latency mode of this pipeline were the
first to identify the second gravitational wave event detected,
GW151226~\cite{abbott2016gw151226}. The pipeline can also operate in an
``offline" configuration that can be used to process archival
gravitational-wave data with additional background statistics and data quality
information. The offline configuration was used in the detection of GW150914,
LVT151012~\cite{abbott2016gw150914}, and GW151226~\cite{abbott2016gw151226}.

The \pipeline{} expands on the parameter space covered by previous low-latency
searches~\cite{abadie2012first, klimenko2008cwb, lynch2015gwb,
TheLIGOScientific:2016uux}. In addition, it extends many of the techniques used
in prior searches for compact binary coalescences~\cite{buskulic2010very,
babak2013searching} to operate in a fully parallel, stream-based mode that
allows for the identification of candidate gravitational-wave events within
seconds of recording the data.  The key differences include: (1)
time-domain~\cite{cannon2012toward} rather than
frequency-domain~\cite{allen2012findchirp} matched filtering, (2) time-domain
rather than frequency-domain~\cite{allen2005chi} signal consistency tests to
reject non-stationary noise transients, (3) a multidimensional likelihood ratio
ranking statistic to robustly identify gravitational-wave candidates in a way
that automatically adjusts to the properties of the
noise~\cite{cannon2015likelihood}, and (4) a background estimation technique
that relies on tracking noise distributions to allow rapid evaluation of
significance of identified candidates~\cite{cannon2013method}.  For a
discussion of performance differences between time-domain and frequency-domain
matched filtering, the reader is referred to Ref.~\cite{cannon2012toward}.

This paper is organized as follows: In Sec.~\ref{sec:initial}, we discuss
inputs to the low-latency and offline analyses, the online acquisition of data,
measurement of the \ac{PSD}, and whitening and conditioning of the data for
matched filtering.  We also present the basic offline and low-latency workflows
in Fig.~\ref{f:offline} and Fig.~\ref{f:online}, respectively. In
Sec.~\ref{sec:ident}, we discuss the matched-filter algorithm and our procedure
for producing a list of ranked candidate events.  In Sec.~\ref{sec:final}, we
explain the significance calculation for identified candidate events and the
procedure for responding to significant events via alerts to our observing
partners. Differences between the offline and low-latency operation modes will
be highlighted when relevant. 

\input{tikzfigure1.tex}

\input{tikzfigure2.tex}

%% file: tikzfigure1.tex
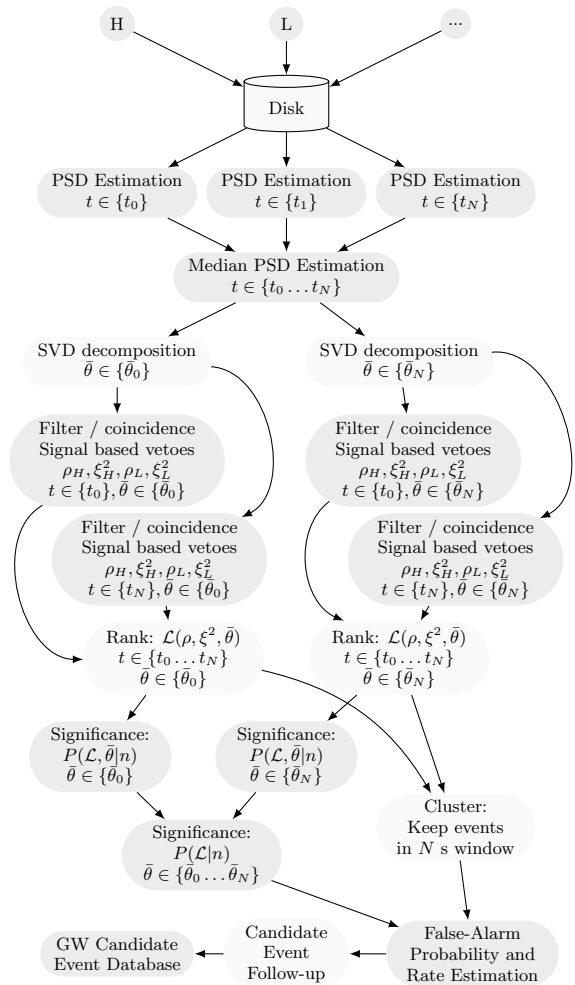
\begin{figure}[h!]
\begin{tikzpicture}[scale=.75, transform shape, >=latex]

	\node[circle, fill=gray!15] (H) at (0, 20) {H};
	\node[circle, fill=gray!15] (L) at (3, 20) {L};
	\node[circle, fill=gray!15] (X) at (6, 20) {...};

	\node[cylinder, fill=gray!5, shape border rotate=90, draw, minimum height=1.0cm, minimum width=1.5cm, shape aspect=.25,] (disk) at (3,18.5) {Disk};

	\node[rounded rectangle, fill=gray!15, align=center] 
		(psd1) at (0, 17) {PSD Estimation\\ $t\in\{t_0\}$};
	\node[rounded rectangle, fill=gray!15, align=center] 
		(psd2) at (3, 17) {PSD Estimation\\ $t\in\{t_1\}$};
	\node[rounded rectangle, fill=gray!15, align=center] 
		(psdN) at (6, 17) {PSD Estimation\\ $t\in\{t_N\}$};
	\node[rounded rectangle, fill=gray!15, align=center] 
		(medpsd) at (3, 15.5) {Median PSD Estimation\\ $t\in\{t_0 \dots t_N\}$};

	\node[rounded rectangle, fill=gray!5, align=center] 
		(svd1) at (0, 14.) {SVD decomposition\\$\params\in\{\params_0 \}$};
	\node[rounded rectangle, fill=gray!5, align=center] 
		(svd2) at (5, 14.) {SVD decomposition\\$\params\in\{\params_{N}\}$};
	
	\node[rounded rectangle, fill=gray!15, align=center] 
		(insp1) at (0, 12.25) {Filter / coincidence\\Signal based vetoes\\ $\rho_H, \xi_H^2, \rho_L, \xi_L^2$\\$t\in\{t_0\}, \params\in\{\params_0\}$};
	\node[rounded rectangle, fill=gray!15, align=center] 
		(insp2) at (0.75, 10.5) {Filter / coincidence\\Signal based vetoes\\ $\rho_H, \xi_H^2, \rho_L, \xi_L^2$\\$t\in\{t_N\}, \params\in\{\params_0\}$};
	\node[rounded rectangle, fill=gray!15, align=center] 
		(insp3) at (5.25, 12.25) {Filter / coincidence\\Signal based vetoes\\ $\rho_H, \xi_H^2, \rho_L, \xi_L^2$\\$t\in\{t_0\}, \params\in\{\params_N\}$};
	\node[rounded rectangle, fill=gray!15, align=center] 
		(insp4) at (6, 10.5) {Filter / coincidence\\Signal based vetoes\\ $\rho_H, \xi_H^2, \rho_L, \xi_L^2$\\$t\in\{t_N\}, \params\in\{\params_N\}$};

	\node[rounded rectangle, fill=gray!5, align=center] 
		(like1) at (1, 8.75) {Rank: $\likehood(\rho,\xi^2,\params)$\\$t\in\{t_0 \dots t_N\}$\\$\params\in\{\params_0 \}$};
	\node[rounded rectangle, fill=gray!5, align=center] 
		(like2) at (5, 8.75) {Rank: $\likehood(\rho,\xi^2,\params)$\\$t\in\{t_0 \dots t_N\}$\\$\params\in\{\params_N\}$};

	\node[rounded rectangle, fill=gray!5, align=center] 
		(cluster) at (6, 5.75) {Cluster:\\Keep events\\in $N$ s window};

	\node[rounded rectangle, fill=gray!15, align=center] 
		(bg1) at (-.3, 7) {Significance:\\$P(\likehood,\params|n)$\\$\params \in \{\params_0\}$};
	\node[rounded rectangle, fill=gray!15, align=center] 
		(bg2) at (3, 7) {Significance:\\$P(\likehood,\params|n)$\\$\params \in \{\params_{N}\}$};
	\node[rounded rectangle, fill=gray!15, align=center] 
		(bgN) at (1.5, 5.25) {Significance:\\$P(\likehood|n)$\\$\params \in \{\params_0\dots\params_N\}$};
	\node[rounded rectangle, fill=gray!15, align=center] 
		(far) at (6.3, 3.5) {False-Alarm \\Probability and \\Rate Estimation};
	\node[rounded rectangle, fill=gray!5, align=center] 
		(fu) at (3, 3.5) {Candidate\\Event\\Follow-up};
	\node[rounded rectangle, fill=gray!15, align=center] 
		(gracedb) at (0, 3.5) {GW Candidate\\Event Database};

	\draw [->] (H) -> (disk);
	\draw [->] (L) -> (disk);
	\draw [->] (X) -> (disk);
	\draw [->] (disk) -> (psd1);
	\draw [->] (disk) -> (psd2);
	\draw [->] (disk) -> (psdN);
	\draw [->] (psd1) -> (medpsd);
	\draw [->] (psd2) -> (medpsd);
	\draw [->] (psdN) -> (medpsd);
	\draw [->] (medpsd) -> (svd1);
	\draw [->] (medpsd) -> (svd2);
	\draw [->] (svd1) -> (insp1);
	\draw [->] (svd1) edge[bend left=75] (insp2);
	\draw [->] (svd2) -> (insp3);
	\draw [->] (svd2) edge[bend left=80] (insp4);
	\draw [->] (insp1) edge[bend right=75] (like1);
	\draw [->] (insp2) ->  (like1);
	\draw [->] (insp3) edge[bend right=60] (like2);
	\draw [->] (insp4) ->  (like2);
	\draw [->] (like1) edge[bend left=23] (cluster);
	\draw [->] (like2) -> (cluster);
	\draw [->] (like1) -> (bg1);
	\draw [->] (like2) -> (bg2);
	\draw [->] (bg1) -> (bgN);
	\draw [->] (bg2) -> (bgN);
	\draw [->] (cluster) -> (far);
	\draw [->] (bgN) -> (far);
	\draw [->] (far) -> (fu);
	\draw [->] (fu) -> (gracedb);

\end{tikzpicture}
\caption{\label{f:offline}
Diagram of the offline search mode of the \gstlal{} based inspiral pipeline.
First, data is transferred from each observatory (H,L,$\dots$) to a central
computing cluster (Sec.~\ref{ss:data}). Next, data is read from disk and the
\ac{PSD} is estimated (Sec.~\ref{ss:psd}) in chunks of time $t_0, t_1, \dots, t_N$
for each observatory. The median over the entire analysis time of each
observatory \ac{PSD} estimate is computed. The input template bank, which is
generated upstream of the analysis, is split into regions of similar parameters
$\params_0, \params_1, \dots, \params_N$ (Sec.~\ref{ss:bank}) and then
decomposed into a set of orthonormal filters weighted by the median \ac{PSD} for
each observatory.  The data is filtered to produce a series of triggers
characterized by \ac{SNR}, $\rho$, signal consistency check, $\xi^2$
(Secs.~\ref{ss:matched}, \ref{ss:triggering}, and \ref{ss:sigveto}), and
coalescence time.  Coincident triggers between detectors are identified and
promoted to the status of events (Sec.~\ref{ss:coinc}). Events are ranked
according to their relative probability of arising from signal versus noise
(Sec.~\ref{ss:rank}).  The data is then reduced to the most highly ranked event
in 8 second windows (Sec.~\ref{ss:cluster}). In parallel, triggers not found in
coincidence are used to construct the probability of obtaining a given event
from noise, $P(\Lambda|n)$. Finally, the event significance and False-Alarm
Rate are estimated (Sec.~\ref{ss:sig}). Note that the arrows drawn between
nodes in this diagram do not imply the output of one node is the input of the
next node, they simply indicate the order in which tasks are performed. 
}
\end{figure}

%% file: tikzfigure2.tex
\begin{figure*}
\begin{tikzpicture}[scale=.9, transform shape, >=latex]
	\node[circle, fill=gray!15] (H) at (0, 4) {H};
	\node[circle, fill=gray!15] (L) at (0, 2) {L};
	\node[circle, fill=gray!15] (X) at (0, 0) {...};
	\node[circle, fill=gray!15] (F) at (18, 0.5) {F};
	\node[circle, fill=gray!15] (S) at (18, 2.5) {S};
	\node[rounded rectangle, fill=gray!5, align=center] 
		(data) at (3, 3) {Data\\Broadcast};
	\node[rounded rectangle, fill=gray!15, align=center] 
		(inj) at (3, 1.5) {Injection\\Broadcast};
	\node[rounded rectangle, fill=gray!5, align=center] 
		(back) at (3, 0) {Background\\Estimation};
	\node[rounded rectangle, fill=gray!15, align=center] 
		(node1) at (7, 3) {Filter / coincidence\\Signal based vetoes / Rank\\ $\rho_H, \xi_H^2, \rho_L, \xi_L^2, \params\in\{\params_0\}$};
	\node[rounded rectangle, fill=gray!15, align=center] 
		(node2) at (7, 1.5) {Filter / coincidence\\Signal based vetoes / Rank\\ $\rho_H, \xi_H^2, \rho_L, \xi_L^2, \params\in\{\params_1\}$};
	\node[rounded rectangle, fill=gray!15, align=center] 
		(nodeN) at (7, 0) {Filter / coincidence\\Signal based vetoes / Rank\\ $\rho_H, \xi_H^2, \rho_L, \xi_L^2, \params\in\{\params_N\}$};
	\node[rounded rectangle, fill=gray!15, align=center] 
		(gracedb) at (12, 1.5) {Gravitational-Wave\\Candidate Event\\Database};
	\node[rounded rectangle, fill=gray!5, align=center] 
		(bayestar) at (11, 3.0) {BAYESTAR\\Rapid Sky\\Localization};
	\node[rounded rectangle, fill=gray!5, align=center] 
		(raven) at (14, 3.0) {RAVEN GRB\\Coincidence};
	\node[rounded rectangle, fill=gray!5, align=center] 
		(lalinference) at (11, 0.0) {lal\_inference\\Parameter\\Estimation};
	\node[rounded rectangle, fill=gray!5, align=center] 
		(fu) at (14, 0.0) {DQ and Candidate\\Follow-up};
	\node[rounded rectangle, fill=gray!5, align=center] 
		(approval) at (15.5, 1.5) {Approval\\Processor};
	\node[rounded rectangle, fill=gray!15, align=center] 
		(gcn) at (18, 1.5) {GCN};
	\draw [->] (H) -> (data);
	\draw [->] (L) -> (data);
	\draw [->] (X) -> (data);
	\draw [->] (data) -> (inj);
	\draw [->] (data) -> (node1);
	\draw [->] (data) -> (node2);
	\draw [->] (data) edge[bend right=15] (nodeN);
	\draw [->] (inj) -> (node1);
	\draw [->] (inj) -> (node2);
	\draw [->] (inj) -> (nodeN);
	\draw [<->] (back) edge[bend left=15] (node1);
	\draw [<->] (back) -> (node2);
	\draw [<->] (back) -> (nodeN);
	\draw [->] (node1) -> (gracedb);
	\draw [->] (node2) -> (gracedb);
	\draw [->] (nodeN) -> (gracedb);
	\draw [<->] (gracedb) -> (bayestar);
	\draw [<->] (gracedb) -> (lalinference);
	\draw [<->] (gracedb) -> (raven);
	\draw [<->] (gracedb) -> (fu);
	\draw [->] (gracedb) -> (approval);
	\draw [->] (approval) -> (gcn);
	\draw [->] (S) -> (gracedb);
	\draw [->] (F) -> (gracedb);

	\draw [thick,decorate,decoration={brace,amplitude=10pt,mirror},xshift=0.4pt,yshift=-0.4pt](-0.5,-0.75) -- (1.5,-0.75) node[black,midway,yshift=-0.8cm,align=center] {\footnotesize GW Observatories\\$\mathcal{O}(10)$s latency};

	\draw [thick,decorate,decoration={brace,amplitude=10pt,mirror},xshift=0.4pt,yshift=-0.4pt](2,-0.75) -- (9.25,-0.75) node[black,midway,yshift=-0.8cm,align=center] {\footnotesize Central Computing Cluster\\$\mathcal{O}(30)$s latency};

	\draw [thick,decorate,decoration={brace,amplitude=10pt,mirror},xshift=0.4pt,yshift=-0.4pt](9.75,-0.75) -- (16.5,-0.75) node[black,midway,yshift=-0.8cm,align=center] {\footnotesize Resources spread\\over the LIGO Data Grid};

	\draw [thick,decorate,decoration={brace,amplitude=10pt,mirror},xshift=0.4pt,yshift=-0.4pt](17,-0.75) -- (19,-0.75) node[black,midway,yshift=-0.8cm,align=center] {\footnotesize Outside world};
\end{tikzpicture}
\caption{\label{f:online}
Diagram of the low-latency search mode of the \gstlal{} based inspiral
pipeline.  First, data is received over a network connection from each
observatory to a data broadcaster in a central computing facility. The data is
then broadcast over the entire cluster with an efficient multicast protocol.
The online analysis uses precomputed bank decompositions for each observatory
from reference \acp{PSD} as input to jobs that combine the filtering, vetoing,
coincidence, ranking, and significance estimation steps from the offline
pipeline. Unlike the offline case, the online analysis work-flow can not be
described as a directed acyclic graph, and in fact, data from each filtering
job is exchanged bi-directionally and asynchronously to a process that
constantly evaluates the global background estimates for the entire analysis.
Events that are identified by any one filtering job, and subsequently pass a
predetermined significance threshold, are sent to the Gravitational-Wave
Candidate Event Database (GraceDB) \cite{gracedb} within a matter of seconds of
the data being recorded at the observatories. 
}
\end{figure*}
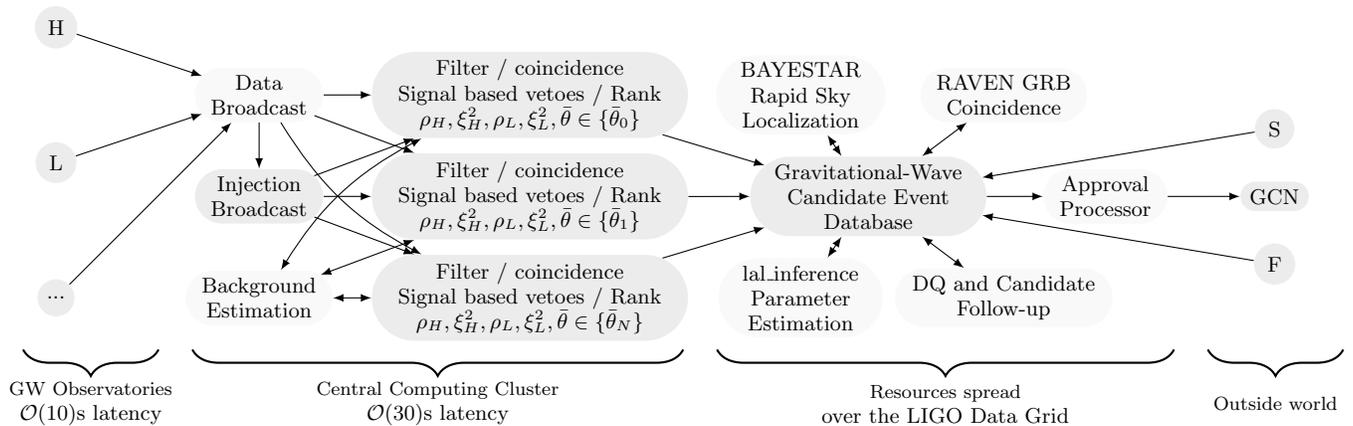

%% file: sec2intro.tex
Matched filtering algorithms for compact binary mergers have traditionally
filtered the data $d(t)$ against a set of complex template waveforms
$\{h^{\mathrm{c}}_i(t)\}$ in the frequency domain using the relation
	 \begin{align}
    \label{eq:complexsnr}
    z_i(t) &= x_i(t) + iy_i(t) = 4 \int_0^\infty \mathrm{d}f \frac{\widetilde{h}^{\mathrm{c}*}_i(f) \widetilde{d}(f)}{S_n(f)} e^{2 \pi i f t},
  \end{align} 
where $z_i(t)$ is the complex \ac{SNR} using the $i$th template, $x_i(t)$ is
the matched filter response to a gravitational wave signal with orbital
coalescence phase $\phi_0$ (the real part of the template in the time domain),
$y_i(t)$ is the matched filter response to the same signal with orbital
coalescence phase $\phi_0 + \pi/4$ (the imaginary part of the template in the
time domain), and $S_n(f)$ is the single-sided noise \ac{PSD}.  The templates
are normalized such that
 \begin{align}
	1 &= 4 \int_0^\infty \mathrm{d}f \frac{\left| \widetilde{h}^{\mathrm{c}}_i(f) \right|^2}{S_n(f)}. \label{eq:complexnormalization}
\end{align} 
Defining the \ac{SNR}, $\rho(t)$, as the modulus of the complex \ac{SNR},
$z(t)$, allows one to search efficiently over the unknown coalescence phase,
while its expression in the frequency domain allows one to efficiently
implement matched filtering using FFT routines.

The \gstlal{}-based inspiral pipeline, however, performs matched filtering in
the time domain with real templates $\{h_i(t)\}$. The matched filter output is
thus the real-valued $x_i(t)$ instead of the complex-valued $z_i(t)$. We can
recast Eq.~\eqref{eq:complexsnr} in the time domain using the convolution
theorem, which gives
\begin{subequations}
\label{eq:realmf}
 \begin{align}
	x_i(t) &= 2 \int_{-\infty}^\infty \mathrm{d}f \frac{\widetilde{h}_i^*(f) \widetilde{d}(f)}{S_n(|f|)} e^{2 \pi i f t} \label{eq:gstlalfreqmf} \\
	&= 2 \int_{-\infty}^\infty \mathrm{d} \tau \, \hat{h}_i(\tau) \hat{d}(t+\tau), \label{eq:gstlaltimemf}
\end{align} 
\end{subequations}
where
 \begin{align}
        \hat{d} (\tau) &= \int_{-\infty}^\infty \mathrm{d}f \frac{\widetilde{d}(f) }{\sqrt{S_n(|f|)}} e^{2 \pi i f \tau}
\end{align} 
is the whitened data; the whitened template $\hat{h}_i(\tau)$ is defined
similarly. As a consequence of using real templates, Eq.~\eqref{eq:realmf}
returns the matched filter response to a single coalescence phase, while
Eq.~\eqref{eq:complexsnr} returns the response to two phases. Thus,
Eq.~\eqref{eq:realmf} must be evaluated a second time using the template
corresponding to the $\pi/4$-shifted phase in order to compute the \ac{SNR}.
To account for using twice as many templates, the template index, $i$, used on
real templates is related to the index used on complex templates via
\begin{subequations}
 \begin{align}
	h_{2i}(t) &= \mathrm{Re} \left[ h_i^\mathrm{c}(t) \right], \\
	h_{2i+1}(t) &= \mathrm{Im} \left[ h_i^\mathrm{c}(t) \right] .
\end{align} 
\end{subequations}
A different template normalization is also used, specifically
 \begin{align}
	1 &= 4 \int_0^\infty \mathrm{d}f \frac{\left| \widetilde{h}_i(f) \right|^2}{S_n(f)}. \label{eq:normalization}
\end{align} 

In this section, we discuss the inputs to the time-domain matched filtering
calculation expressed in Eq.~\eqref{eq:realmf}. We begin by discussing the
low-latency distribution of the data itself. We then describe our method for
estimating the \ac{PSD}, which we use to construct the whitened data
$\hat{d}(t)$ and whitened templates $\hat{h}_i(t)$. In describing our
construction of the whitened data stream, we also describe the removal of loud
noise transients and dealing with data dropouts in the low-latency
broadcast. Finally, we describe the construction of the whitened template
filters, which involves a number of computational enhancements to reduce the
cost of filtering in the time domain.

%% file: data.tex
Gravitational-wave strain data acquired at the LIGO sites is digitized at a
sample rate of $\unit{16384}{\hertz}$ and bundled into IGWD frames, a custom
LIGO file format described in Ref.~\cite{FrameFormat}, on a four-second
cadence. Information about the state of the instrument and data quality are
distilled from a host of auxiliary environmental and
instrumental-control-system channels into a single channel, referred to as the
state-vector channel. The four-second frames containing the gravitational-wave
strain and state-vector channels are delivered for low-latency processing at
computing clusters across the LIGO Data Grid within $\sim 12$ seconds of the
data being acquired.  

Searches for compact binary coalescences require using hundreds or thousands of
compute nodes in parallel to process all the possible template waveforms.
Low-latency data must be made available to all of these nodes as soon as it
arrives, thus an efficient multicast protocol is used to broadcast the data in
low latency to the entire cluster.  The nature of the low-latency transmission
causes some small data loss within the tolerances acceptable to the pipeline,
with efforts underway to reduce these losses.

%% file: psd.tex
Abstractly, we define the (one-sided) noise power spectral density $S_n(f)$ as
 \begin{align}
	\langle \widetilde{n}(f) \widetilde{n}^*(f') \rangle = \frac{1}{2} S_n(f) \delta(f - f'), f > 0\label{eq:sspsd}
\end{align} 
where $\langle \cdots \rangle$ denotes an ensemble average over realizations of
the detector noise $n(t)$, which is assumed to be stationary and Gaussian. In
practice, we cannot use Eq.~\eqref{eq:sspsd} to calculate the \ac{PSD} for a
variety of reasons. To begin with, our knowledge of the detector noise comes
exclusively from the observed data, which may contain signal in addition to
noise. Furthermore, real data may contain brief departures from stationarity
(commonly called ``glitches''), which we do not want to contribute to the
\ac{PSD} estimate. Finally, the \ac{PSD} can drift slowly over time scales
shorter than the duration of a typical detector lock segment, and we want to
track these changes. For low-latency applications, we also require a \ac{PSD}
estimate that converges quickly using only data in the past, so that we obtain
an accurate estimate of the \ac{PSD} as soon as possible after the data begin
to flow. In this subsection, we discuss the \ac{PSD} estimation algorithm and
how the result is used to whiten the data and template bank. We also present
the results of a study done on the convergence of an estimated \ac{PSD} to its
known spectrum.

\subsubsection{Estimation and Whitening}
We use a median and a running geometric mean to meet these requirements for
each analyzed segment of data. The median estimate operates on medium time
scales and is robust against shorter time-scale fluctuations in the noise,
while the running geometric mean tracks longer time-scale changes in the
\ac{PSD}, averaging the \ac{PSD} estimates with the most recent estimates
weighted more strongly. The time scales of the median and geometric mean are
set, respectively, by the tunable parameters $n_\mathrm{med}$ and
$n_\mathrm{avg}$. 

The \ac{PSD} calculation begins by partitioning the strain time series into
blocks of length $N$ points with each block overlapping the previous by $N/2 +
Z$ points, where $N$ and $Z$ are even-valued integers.  Each block of data,
denoted $d_j [k]$, is windowed and Fourier transformed,
 \begin{equation}
   \widetilde{d}_j[\ell] = \sqrt{\frac{N}{\sum_{k=0}^{N-1} w[k]^2}} \Delta t \sum_{k=0}^{N-1} d_j[k] w[k] e^{-2 \pi i \ell k / N}, \label{eq:datafourier}
\end{equation} 
where $k \in [0,N-1]$ is the time index, $\ell \in [0,\frac{N}{2}]$ is the
frequency index, $\Delta t$ is the time sample step, and
 \begin{equation}
w[k]
   = \begin{cases}
   0, & 0 \leq k < Z \\
   \sin^{2} \frac{\pi (k - Z)}{N - 2 Z}, & Z \leq k < N - Z \\
   0, & N - Z \leq k \leq N-1 \\
   \end{cases}
\end{equation} 
is a zero-padded Hann window function. The DC and Nyquist terms of
Eq.~\ref{eq:datafourier}, $\widetilde{d}[0]$ and $\widetilde{d}[N/2]$, are set
to zero, and the zero-padded Hann window is defined such that the sequence of
overlapping window functions sum to unity everywhere. The squared magnitude of
Eq.~\eqref{eq:datafourier} is proportional to the instantaneous \ac{PSD} and
has a frequency resolution of $\Delta f = \frac{1}{N \Delta t}$.

The median of the most recent $n_{\mathrm{med}}$ instances of the instantaneous
\ac{PSD}, $S^\mathrm{med}_j[\ell]$, is determined for each frequency bin $\ell$.
Mathematically,
 \begin{equation} 
	S^\mathrm{med}_j[\ell] = \mathrm{median} \{~ 2 \Delta f |\widetilde{d}_k[\ell]|^2~
	\}_{k=j-n_{\mathrm{med}}}^{k=j}.
\end{equation} 
The median is relatively insensitive to short time-scale fluctuations, which
must occur over a time scale of $\frac{1}{2} n_{\mathrm{med}} (N/2 -Z) \Delta
t$ in order to affect the median. 

The median is used to estimate the geometric mean of the last
$n_{\mathrm{med}}$ samples for each frequency bin. Assuming the noise is a
stationary, Gaussian process allows us to assume that the measured frequency
bins of the estimated \ac{PSD} are $\chi^2$-distributed random variables. The
geometric mean of a $\chi^2$-distributed random variable is equal to the median
divided by a proportionality constant $\beta$. The logarithm of the running
geometric mean of median estimated \acp{PSD}, $\log S_j[\ell]$, is computed
from one part $\log S_j^{\mathrm{med}}[\ell]/\beta$ and $(n_\mathrm{avg} - 1)$
parts $\log S_{j-1}[\ell]$. Mathematically,
\begin{equation}
S_{j}[\ell]
   = \exp \left[
   \frac{n_{\mathrm{avg}} - 1}{n_{\mathrm{avg}}} \log S_{j-1}[\ell] +   \frac{1}{n_{\mathrm{avg}}} \log \frac{S^{\mathrm{med}}_{j}[\ell]}{\beta}   \right]. \label{eq:geomean}
\end{equation}
Changes to the \ac{PSD} must occur over a time scale of at least
$n_{\mathrm{avg}} (N/2 - Z) \Delta t$ to be fully accounted for by
Eq.~\eqref{eq:geomean}.

To whiten the data and the templates, the arithmetic mean is estimated from the
geometric mean. The arithmetic mean of a $\chi^2$-distributed random variable
is equal to the geometric mean multiplied by $\exp (\gamma)$, where $\gamma$ is
Euler's constant. If the noise assumptions are violated, then the true
arithmetic mean of the spectrum will differ from the measured spectrum by some
unknown factor. This estimated arithmetic mean is referred to as $S_n(f)$ in
the continuum limit (see e.g.  Eq.~\ref{eq:sspsd}).

The low-latency operating mode must whiten the data and update the running
geometric mean of the \ac{PSD} at the same time. The whitening process is done
after the running geometric mean has been updated and is performed by dividing
each frequency bin of Eq.~\ref{eq:datafourier} by the square root of the
corresponding frequency bin in the estimated arithmetic mean of the \ac{PSD}.
Mathematically,
 \begin{align}
	\widetilde{\hat{d}}_j  [\ell] &= \frac{\widetilde{d}_j [\ell]}{\sqrt{S_j [\ell] \exp ( \gamma) }}, \\
	\hat{d}_j [k] &= 2 \Delta t \sqrt{\sum_{m=0}^{N-1} w[m]^2} \Delta f \sum_{\ell=0}^{N/2} \widetilde{\hat{d}}_j[\ell] e^{2 \pi i \ell k / N}.
\end{align} 
The extra terms in the inverse Fourier transform are necessary for unity
variance.  

The low-latency analysis typically uses $N=f_s (\unit{8}{\second})$ and $Z=f_s
(\unit{2}{\second})$, where $f_s = 1 / \Delta t$ is the sampling frequency,
resulting in $\unit{1/4}{\hertz}$ frequency resolution. This introduces four
seconds of latency into the analysis.  Unlike the low-latency case, the offline
analysis begins with a known list of data segments. The \ac{PSD} of each
segment is estimated using $N=f_s (\unit{32}{\second})$ and $Z=0$; the final
result of the running average is written to disk as a ``reference \ac{PSD}.''
The median of the reference \acp{PSD} is used to whiten the template bank
before matched filtering.  However, the data segments are whitened in a
procedure similar to the low-latency analysis, using $N = f_s
(\unit{32}{\second})$ and $Z = f_s (\unit{8}{\second})$.  At the time of
writing, the typical values used are $n_\mathrm{med} = 7$ and $n_\mathrm{avg} =
64$ for both modes of operation. The only procedural difference between the
offline and low-latency whitening steps is that the offline analysis seeds the
running average with the segment's reference \ac{PSD}.

\subsubsection{Convergence}
\begin{figure}[h]
\includegraphics[width=0.95\columnwidth]{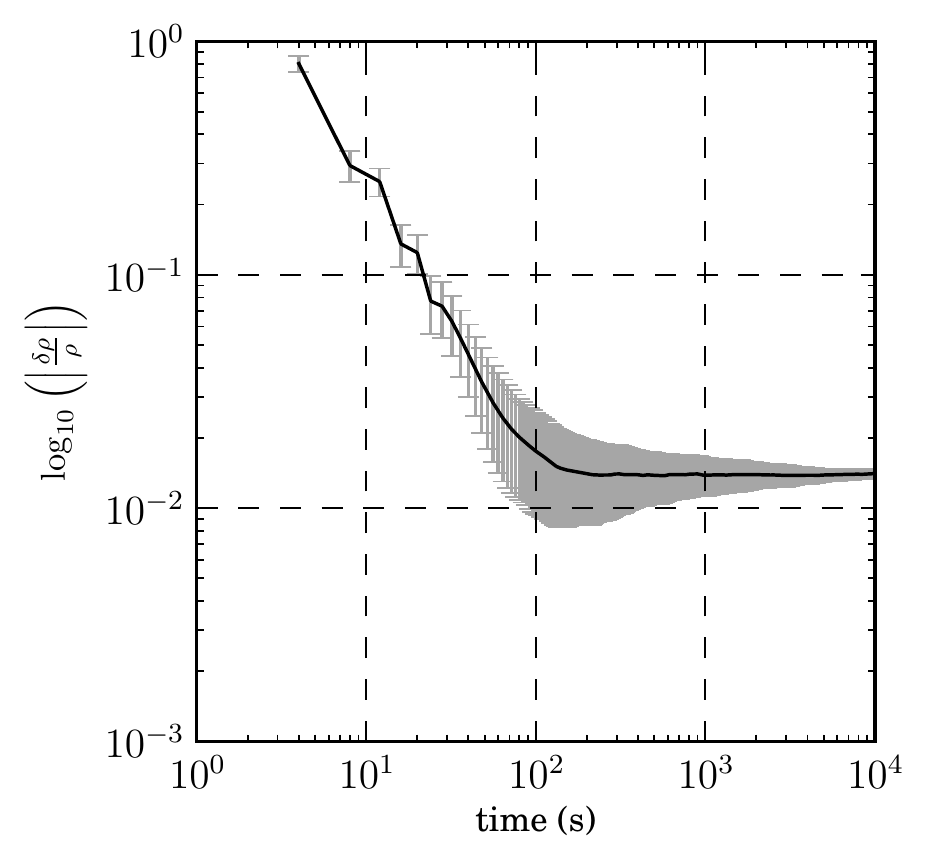}
\caption{\label{psdconvergence} \ac{PSD} convergence properties.
Estimating the \ac{PSD} is a critical part of ensuring that events are
detected and assigned the appropriate significance.  This figure
illustrates the convergence properties of the \ac{PSD} estimation in terms
of the impact on SNR.  Within $\unit{20}{\second}$ the \ac{PSD} will have
an $\order(10\%)$ impact on SNR, and within $\unit{200}{\second}$ the impact
drops to $\order(1\%)$, where it remains.}
\end{figure}

For low-latency applications, we require a \ac{PSD} estimate that converges
quickly using only data in the past, so that we obtain an accurate estimate of
the \ac{PSD} as soon as possible after the data begin to flow. To quantify the
convergence, we create noise with a known power spectrum and compute a quantity
that is proportional to the expected \ac{SNR} for a given \ac{PSD} in the
absence of noise (commonly referred to as the `optimal \ac{SNR}').  In the
stationary phase approximation, binary waveforms in the frequency domain,
$h(f)$, are proportional to $f^{-7/6}$~\cite{droz1999gravitational}, thus
	 \begin{align}
		\rho &\propto \int_{f_1}^{f_2} \mathrm{d} f \frac{f^{-7/3}}{S_n(f)}.
		\label{eq:psdrho}
	\end{align} 
We choose $f_1 = \unit{10}{\hertz}$ and $f_2 = \unit{2048}{\hertz}$.
Specifically, we compare the quantity computed using the \textit{measured} spectrum $S_n(f)$, which we
denote simply as $\rho$, to the SNR computed from the \textit{known} spectrum $\hat{S}_n(f)$, 
which we denote as $\hat{\rho}$.  Fig.~\ref{psdconvergence} shows the
fractional change of $\rho$ with respect to $\hat{\rho}$ as a function of time,
	 \begin{align}\label{frac_rho}
		\frac{\delta\rho}{\rho}(t) = \frac{\rho(t)-\hat{\rho}}{\rho(t)}.
	\end{align} 
We find that convergence happens quickly relative to the length of the
data.  The approximation of the true \ac{PSD} does not affect the
measured SNR after tens of seconds.

%% file: datacond.tex
Matched filtering is optimal under the condition that the noise, $n(t)$, is
Gaussian. Our implementation of matched filtering also assumes stationarity
over time-scales at least as long as the compact binary waveform. Although
non-stationarity on long time scales can be handled by tracking the \ac{PSD},
short noise transients, commonly referred to as ``glitches,'' can cause
high-SNR matched filter outputs that mimic signal detections. Glitches are
handled by either removing them from the data or using signal consistency
checks to vet the matched filter output.  Sec.~\ref{ss:sigveto} provides more
detail on the latter.

\begin{figure}[h]
\includegraphics[width=1\columnwidth]{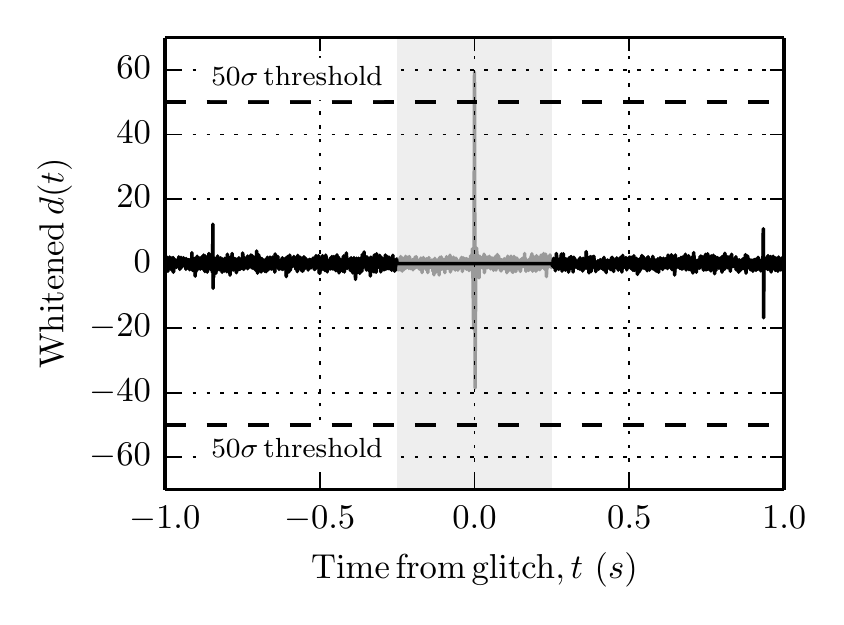}
\caption{Data conditioning.  In this two second block of LIGO S6 data, three
noise transients (``glitches'') are visible.  The glitch at time zero surpassed
the threshold of 50 standard deviations $(\sigma)$, triggering the gate to veto
a $\pm 0.25$ second window around the glitch by replacing the data with zeros
(black).  The gray trace shows what the data looked like prior to gating.
\label{f:data_cond} 
}
\end{figure}

The \gstlal{} based inspiral pipeline removes glitches from the data in two
ways.  In some cases, the matched filter outputs of glitches have considerably
higher amplitude than any expected output from a compact binary signal and can
thus be safely removed from the data through a process called gating. Once the
data has been whitened, it has unit variance.  If a momentary excursion greater
than some number of standard deviations, $\sigma$, is observed in the whitened
data, then the gating process zeros the excursion in the whitened data with a
$\unit{0.25}{\second}$ padding on each side. An example of this is shown in
Fig.~\ref{f:data_cond}. When gating the strain data, care must be taken to
choose a threshold that will not discard real gravitational wave signals. The
threshold is chosen by testing with simulated gravitational wave signals.

\begin{figure}[h]
\includegraphics[width=1\columnwidth]{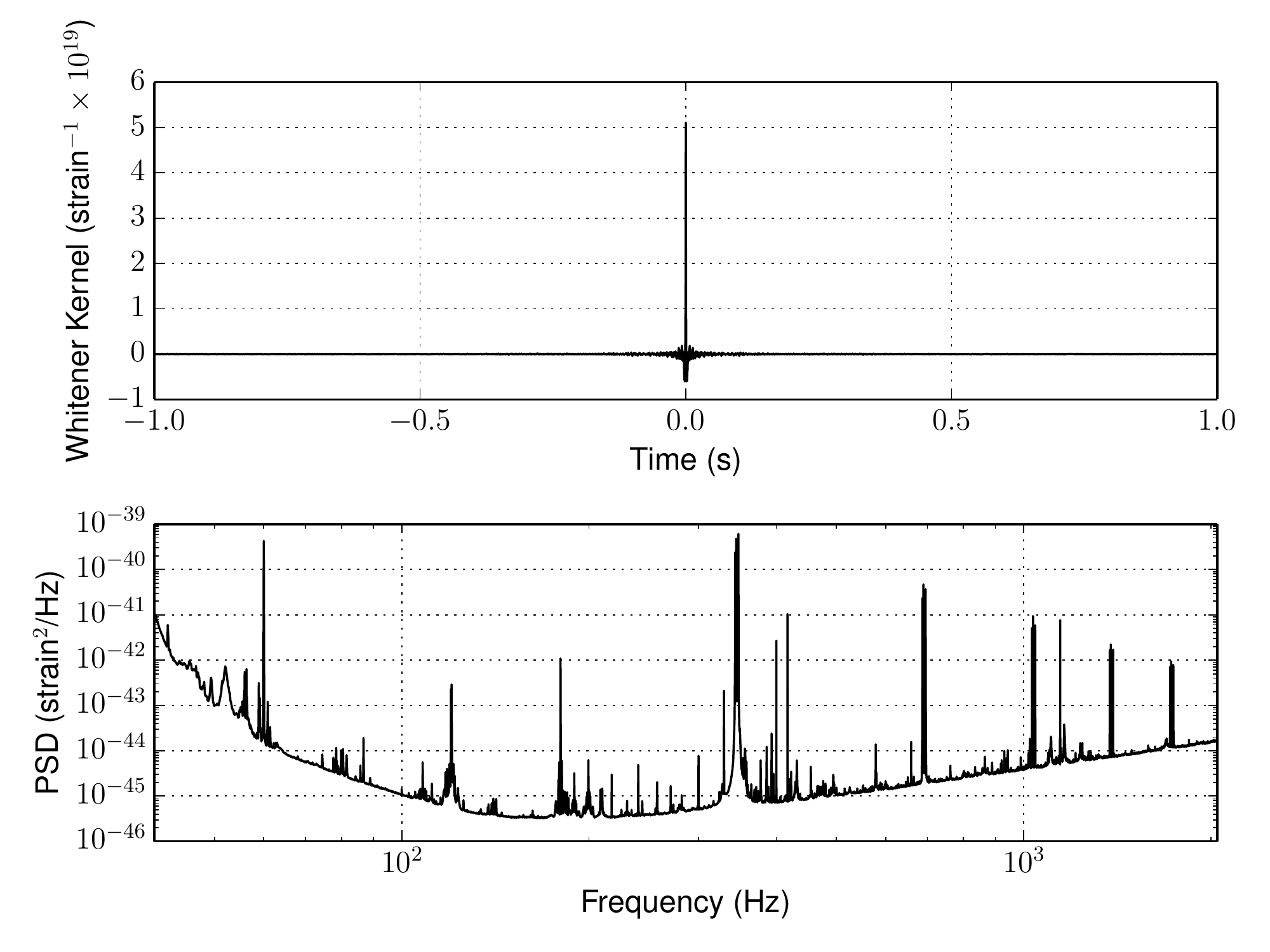}
\caption{Top: Time domain representation of the whitening filter computed from
a \ac{PSD} estimated in an analysis of one week of S6 data.  $\sim 0.98$ of the
filters square magnitude is contained enclosed within $\pm
\unit{10}{\milli\second}$ of the peak. Bottom: The \ac{PSD} used to compute the
whitening filter.}\label{fig:whitefilter}
\end{figure}

The choice of $\unit{0.25}{\second}$ padding is conservative for LIGO
\acp{PSD}, where the whitening filter in the time domain can be approximated as
a narrow sinc function. An Initial LIGO \ac{PSD} and the time domain
representation of its corresponding whitening filter, estimated from data taken
during Initial LIGO's sixth science run
\cite{abadie2012search_a,abadie2012search_b} (referred to as S6), are shown in
Fig.~\ref{fig:whitefilter}.  $\sim 0.98$ whitening filter's square magnitude is
contained within $\pm \unit{10}{\milli\second}$ of the filter's peak, thus we
expect no significant spectral leakage when gating glitches with
$\unit{0.25}{\second}$ padding.

In many cases, auxiliary information is available through environmental
and instrumental monitors that can ascertain times of clear coupling between
local transient noise sources~\cite{slutsky2010methods, christensen2010ligo};
in cases where data quality is known to be poor, vetoes are applied after the
strain data is whitened. Since whitened data is, by definition, uncorrelated
between adjacent samples for stationary Gaussian processes, vetoes are applied
by simply replacing the whitened data during vetoed times with zeros.


%% file: bank.tex
In order to detect any compact binaries within a region of the mass parameter
space, we filter the data against a bank of template signals. As the true
binary parameter space is continuous, actual signals may not exactly match any
one template from the bank; such signals incur a loss of \ac{SNR}. The
parameters of the templates in the bank are chosen to minimize this loss of SNR
using as few templates as possible~\cite{owen1996search, owen1999matched,
apostolatos1995search}.  Techniques for efficiently covering the binary
parameter space with templates have been extensively
developed~\cite{cokelaer2007gravitational, abbott2008search,
harry2014investigating, babak2008building, harry2009stochastic, manca2010cover,
privitera2014improving}.  We assume here that such a template bank has already
been constructed, and describe how the bank is decomposed to more efficiently
filter the data.

The standard methods for template bank construction naturally lead to banks of
highly redundant templates. In the frequency-domain, filtering directly with
the physical templates has the advantage of admitting computationally-efficient
searches over the unknown signal coalescence phase and time; this advantage is
lost in the time-domain. The \pipeline{} therefore does not directly filter the
data against the physical template waveforms themselves. Rather, it employs the
LLOID method~\cite{cannon2012toward} (see also Sec.~\ref{ss:matched}), which
combines \ac{SVD}~\cite{cannon2010singular, cannon2011efficiently,
cannon2012interpolating} with near-critical sampling to construct a reduced set
of orthonormal filters with far fewer samples.

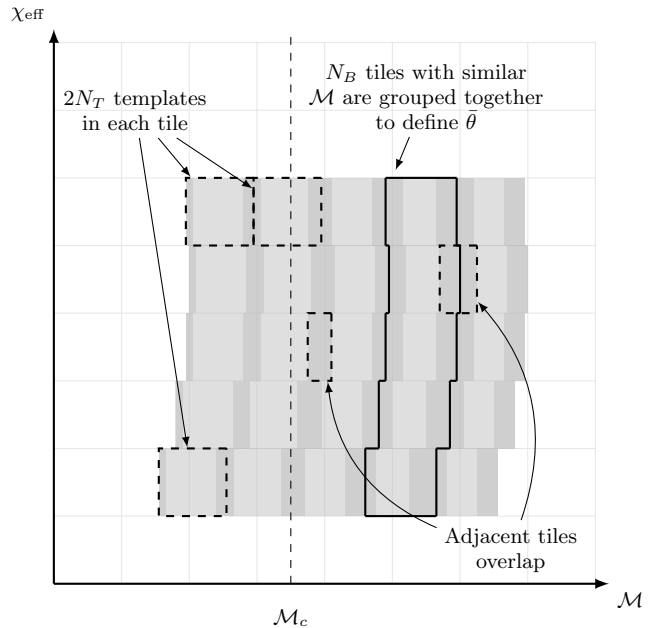
\begin{figure}
\begin{tikzpicture}[scale=.9, transform shape, >=latex]
	\draw[step=1cm,gray!20,very thin] (0,0) grid (8,8);
	\draw[thick,->] (0,0) -- (8.2,0) node[anchor=north west] {$\mchirp$};
	\draw[thick,->] (0,0) -- (0,8.2) node[anchor=south east] {$\chi_{\mathrm{eff}}$};

	\foreach \x in {2,...,2}
		\foreach \y in {-3,...,1}
		{
			{
				\fill[gray!50!white,opacity=0.5] (\x -.05*\y*\y, \y+4) 
					rectangle (\x + 1 + 0.05*\x -.05*\y*\y, \y+5);
				\fill[gray!50!white,opacity=0.5] (\x -.05*\y*\y, \y+4) 
					rectangle (\x -.05*\y*\y +0.05*\x, \y+5);
			}
		}
	\foreach \x in {3,...,5}
		\foreach \y in {-3,...,1}
		{
			{
				\fill[gray!50!white,opacity=0.5] (\x -.05*\y*\y - 0.05*\x, \y+4) 
					rectangle (\x + 1 + 0.05*\x -.05*\y*\y, \y+5);
			}
		}
	\foreach \x in {6,...,6}
		\foreach \y in {-3,...,1}
		{
			{
				\fill[gray!50!white,opacity=0.5] (\x -.05*\y*\y - 0.05*\x, \y+4) 
					rectangle (\x + 1 -.05*\y*\y, \y+5);
				\fill[gray!50!white,opacity=0.5] (\x + 1 -.05*\y*\y - 0.05*\x, \y+4) 
					rectangle (\x + 1 -.05*\y*\y, \y+5);
			}
		}

	\draw [dashed] (3.5,0) -- (3.5,8.1);
	\node (chirpcut) at (3.5,-0.5) [align=center] {$\mchirp_c$};

	\node (t1) at (2.45, 5.5) [draw,thick,dashed,minimum width=1cm,minimum height=1cm] {};
	\node (t2) at (3.45, 5.5) [draw,thick,dashed,minimum width=1cm,minimum height=1cm] {};
	\node (t3) at (2.05, 1.5) [draw,thick,dashed,minimum width=1cm,minimum height=1cm] {};
	\node (temps) at (1.2, 7) [align=center] {$2\NT$ templates\\in each tile};
	\draw[->] (temps) -> (t1);
	\draw[->] (temps) -> (t2);
	\draw[->] (temps) -> (t3);

	\node (overlap1) at (5.975, 4.5) [draw,thick,dashed,minimum width=0.55cm,minimum height=1cm] {}; 
	\node (overlap2) at (3.925, 3.5) [draw,thick,dashed,minimum width=0.35cm,minimum height=1cm] {}; 
	\node (overlaptext) at (6.75, 0.5) [align=center] {Adjacent tiles\\overlap};
	\draw[->] (overlaptext) edge[bend right] (overlap1);
	\draw[->] (overlaptext) edge[bend left] (overlap2);

	\draw [thick] (4.90,5) -- (4.90,6);
	\draw [thick] (4.90,5) -- (4.95,5);
	\draw [thick] (5.95,5) -- (5.95,6);
	\draw [thick] (4.90,6) -- (5.95,6);

	\draw [thick] (5.95,5) -- (6.0,5);
	\draw [thick] (6.0,4) -- (6.0,5);
	\draw [thick] (5.95,4) -- (6.0,4);
	\draw [thick] (5.95,3) -- (5.95,4);
	\draw [thick] (5.85,3) -- (5.95,3);
	\draw [thick] (5.85,2) -- (5.85,3);
	\draw [thick] (5.65,2) -- (5.85,2);
	\draw [thick] (5.65,1) -- (5.65,2);
	\draw [thick] (4.60,1) -- (5.65,1);

	\draw [thick] (4.95,4) -- (4.95,5);
	\draw [thick] (4.90,4) -- (4.95,4);
	\draw [thick] (4.90,3) -- (4.90,4);
	\draw [thick] (4.80,3) -- (4.90,3);
	\draw [thick] (4.80,2) -- (4.80,3);
	\draw [thick] (4.60,2) -- (4.80,2);
	\draw [thick] (4.60,1) -- (4.60,2);

	\node (bin) at (5,6) {};
	\node (bintext) at (5.5, 7.2) [align=center] {$\NB$ tiles with similar \\$\mchirp$ are grouped together\\ to define $\params$};
	\draw[->] (bintext) -> (bin);

\end{tikzpicture}
\caption{\label{f:bank_illustration}
An illustration of how the physical parameter space is tiled into regions in
which the LLOID decomposition is done. The physical parameter space is
projected onto the $\mchirp, \chi$ plane. Tiles of equal template number, $2\NT$,
are constructed and overlapped in the $\mchirp$ direction by
$\mathcal{O}(10\%)$. Above a specified chirp mass, $\mchirp_c$, waveforms that
use the full inspiral-merger-ringdown description are used.  Below $\mchirp_c$,
waveforms that model only the inspiral phase are used.  Tiles of similar chirp
mass are then grouped together to define a one-dimensional family of similar
parameters, $\params$, used in the evaluation of the likelihood-ratio ranking
statistic (Sec.~\ref{ss:rank}). 
}
\end{figure}

\begin{figure}
\includegraphics[width=0.48\textwidth]{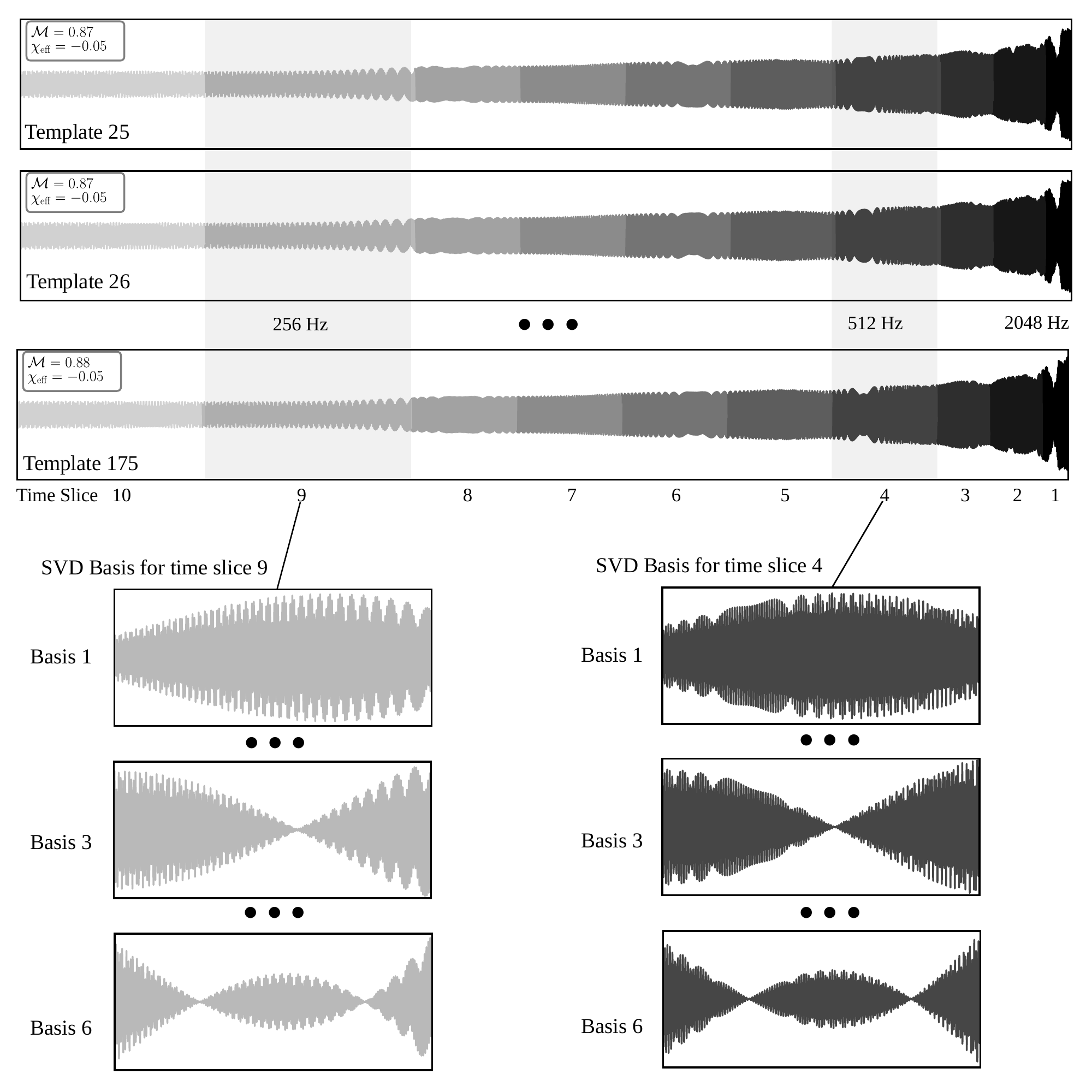}
\caption{\label{f:lloid}
An example of the LLOID decomposition~\cite{cannon2012toward}.  In this
example, $\NT = 195$ binary inspiral waveforms (390 including the two possible
phases) with a chirp mass between 0.87 and 0.88 are first ``whitened'' by
dividing them by a realistic noise amplitude spectral density from \ac{aLIGO}.
The line features in the spectrum are responsible for the amplitude modulation
of the waveforms. The waveforms, which are prepended with zeros when necessary
so that all of the templates in a given decomposition have the same number of
sample points, were decomposed into 30 time slices at sample rates ranging
between 128 Hz and 2048 Hz (only the last 10 slices are shown). A basis filter
set from the waveforms in each time slice was constructed using the
SVD~\cite{cannon2010singular}.  Only 6-10 basis waveforms per slice were needed
to reconstruct both phases of the 195 input waveforms to an accuracy of better
than 99.9\%.
}
\end{figure}

In order to prepare the templates for the LLOID decomposition, the template
bank is first split into partially-overlapping ``split-banks'' of templates
with similar time-frequency evolution based on the template parameters, as
depicted in Fig.~\ref{f:bank_illustration}. Templates corresponding to binary
black hole systems with circular orbits and component spins parallel to the
orbital angular moment can be characterized by the component masses $m_i$ and
the dimensionless spin parameters $\chi_i = \vec{S}_i \cdot \hat{L}/m_i^2$ for
$i=1,2$, where $\vec{S}_i$ are the spin vectors and $\hat{L}$ is the orbital
angular momentum unit vector. Circularized binary neutron star templates with
aligned spins can also be characterized by $m_i$ and $\chi_i$, however not as
accurately due to neutron-star specific effects such as tidal disruption.  The
templates for these systems are binned in a two-dimensional space, first by an
effective spin parameter $\chi_\mathrm{eff}$,
 \begin{align}
\chi_\mathrm{eff} &\equiv \frac{m_1 \chi_1 + m_2 \chi_2}{m_1 + m_2},
\end{align} 
and then by chirp mass $\Mc$,
 \begin{align}
	\Mc &= \frac{(m_1 m_2)^{3/5}}{(m_1 + m_2)^{1/5}}.
\end{align} 
$2\NT$ real templates are placed in each split-bank, where $\NT$ is typically
$\mathcal{O}(100)$. The factor of 2 is a result of using two orthogonal
real-valued templates in place of 1 complex-valued template
(Sec.~\ref{sec:initial}). The input templates in adjacent $\Mc$~bins are
overlapped in order to mitigate boundary effects from the SVD.  Overlapping
regions are clipped after reconstruction such that the output has no redundant
template waveforms\footnotemark.
\footnotetext[1]{Split-banks that contain the lowest and highest $\Mc$
templates in a given $\chi$ bin are padded with duplicate templates from within
the split-bank in order to keep the clipping uniform between split-banks.}
The waveforms are then whitened using reference \acp{PSD}, as described in
Sec.~\ref{ss:psd}, and each split-bank is decomposed via the LLOID method,
described below and in Fig.~\ref{f:lloid}.

Each split-bank is divided into various time slices after prepending the
templates with zeros such that every template has the same number of sample
points; this allows us to efficiently sample different regions of our waveforms
with the appropriate Nyquist frequency instead of over-sampling the
low-frequency regions of the waveform with the sampling frequency required for
the high-frequency regions.  The \ac{SVD} is then performed on each time slice
of each split-bank and truncated such that we retain only the most important
basis waveforms returned by the \ac{SVD} algorithm, as measured by the match
between the original templates and the reconstructed
waveforms~\cite{cannon2012toward}.

In addition to being used for the LLOID decomposition, split-banks are binned
by the lowest chirp mass in each split-bank to construct bins of similar
templates. These are referred to as $\params$ bins, and define a binning of the
likelihood-ratio detection statistic defined in Sec.~\ref{ss:rank}.

%

%% file: eventid.tex
Borrowing the language commonly used in particle experiments, the
\gstlal{}-based inspiral pipeline identifies ``triggers'' from individual
interferometer data streams. Triggers which arrive in coincidence are elevated
to the ``event'' classification and ranked by the likelihood-ratio ranking
statistic.  In this section, we discuss how a list of triggers is generated by
the matched-filtering algorithm and how coincidences are identified and ranked
as events.

%% file: matchedfiltering.tex
As discussed in Sec.~\ref{ss:bank}, groups of templates are partitioned into
time slices as part of the LLOID decomposition~\cite{cannon2012toward}.
Specifically, any split-bank $\mathbf{H}$ can be written as a collection of
matrices $\mathbf{H}^s$,
 \begin{align}
	\mathbf{H} &= \{ \mathbf{H}^s\}, 
\end{align} 
where each $\mathbf{H}^s$ contains time-slice $s$ of all $2\NT$ templates in
the split-bank,
 \begin{align}
	\mathbf{H}^s &= \{\hat{h}_i^s(t) : i \in [0, 2\NT - 1] \}.
\end{align} 
The index $s$ is chosen to be largest at the start of the template waveforms,
decreasing until $s = 0$ for the last slice (as seen in Fig.~\ref{f:lloid}).
Each slice of the split-bank, $\mathbf{H}^s$, is decomposed via the \ac{SVD} to
provide basis functions $u$.  These basis functions can be used to reconstruct
any $\hat{h}_i$ to a predetermined tolerance, i.e.,
 \begin{align}
	\hat{h}_i^s(t) &\approx \sum_{\nu=0}^{N-1} v_{i \nu}^s \sigma_\nu^s u_\nu^s(t), \label{eq:templatesvd}
\end{align} 
where $\mathbf{u}^s=\{u_\nu^s(t)\}$ is a matrix comprised of $N$ basis vectors,
$\mathbf{v}^s=\{v_{i \nu}^s\}$ is a reconstruction matrix, and
$\vec{\sigma}=\{\sigma_\nu^s\}$ is a vector of singular values whose magnitudes
are directly proportional to how important a corresponding basis vector is to
the reconstruction process~\cite{cannon2010singular}. Now instead of evaluating
Eq.~\eqref{eq:gstlaltimemf} $2\NT$ times for each slice of $2\NT$ templates, we
can evaluate
 \begin{align}
	U_\nu^s(t) &= 2 \int_{-\infty}^\infty \mathrm{d} \tau \, u_\nu^s(t) \hat{d}^s(t + \tau) \label{eq:svdmf}
\end{align} 
$N < 2\NT$ times for each slice, where $\hat{d}^s(t)$ is sampled at the same
rate as $u_\nu^s(t)$. The matched filter output time series is calculated for
each time slice $\mathbf{H}^s$, then up-sampled via sinc interpolation and
added to the output of other time slices (in order of decreasing $s$) to obtain
the output of Eq.~\eqref{eq:gstlaltimemf}. The matched filter output
accumulated up through slice $s$ is defined recursively for each template in a
given split-bank as
%
%
 \begin{align}
	x_i^s (t) &= \overbrace{ \left(H^\uparrow x_i^{s+1}\right)(t)}^{\mathrm{Previous\,} x_i} + \underbrace{ \sum_{\nu=0}^{N-1} v_{i\nu}^s \sigma_\nu^s U^s_\nu(t)}_{\mathrm{Current\,}x_i}, \label{eq:recursivesnr}
\end{align} 
where $H^\uparrow$ acts on a time series sampled at $f_{s+1}$ and up-samples it
to $f_s$. Recall that the \gstlal{}-based inspiral pipeline uses two
real-valued templates in place of one complex-valued template
(Sec.~\ref{sec:initial}), thus the computed \ac{SNR} is the quadrature sum of
matched filter outputs from waveforms which differ only in coalescence phase by
$\pi/4$,
 \begin{align}
	\rho_j(t) &= \sqrt{x_{2j}(t)^2 + x_{2j+1}(t)^2}, \, j \in [0,\NT-1].
\end{align} 
Note that there are half as many \acp{SNR} as there are templates, which is a
result of using real-valued templates in place of complex-valued templates
(Sec.~\ref{sec:initial}).  Evaluating Eq.~(\ref{eq:recursivesnr}) saves a
factor of $10^4$ in computational cost over a direct time domain convolution of
the template waveforms for typical \ac{aLIGO} search
parameters~\cite{cannon2012toward}.

%% file: triggering.tex
The raw \ac{SNR} time series, typically sampled at $\unit{2}{\kilo\hertz}$, is
discretized into ``triggers" before being stored. The discretization is done by
maximizing the \ac{SNR} over time in one-second windows and recording the peak
if it crosses a predetermined threshold. With the typical threshold of
$\ac{SNR} = 4$, it is probable to have at least one trigger in every one-second
interval for every template for every detector dataset analyzed.  Although the
number of triggers can easily be hundreds of thousands per second (due to modern
templates banks containing hundreds of thousands of
templates~\cite{abbott2016gw150914}) storing them is a marked improvement over
storing the raw \ac{SNR} time series, which is over three orders of magnitude
more voluminous.  However, we do not discard the raw \ac{SNR} time series
information immediately, because it is needed for the next stage of the
pipeline (Sec.~\ref{ss:sigveto}).  For each trigger, we record the parameters
of the template, the trigger time, the \ac{SNR}, and the coalescence phase. The
trigger time is computed via sub-sample interpolation to nanosecond precision;
while low \ac{SNR} triggers suffer from poor timing resolution, high \ac{SNR}
triggers can be resolved to better resolution than that of the sample rate
\cite{fairhurst2009triangulation, singer2016rapid, singer2014first}.  Triggers
are identified in parallel across each template in a given $\params$ bin.

%% file: autoveto.tex
Detector data often contain glitches that are not removed during the data
conditioning stage (Sec.~\ref{ss:data_cond}). Therefore, ranking triggers
solely by \ac{SNR} is not sufficient to separate noise from transient signals.
Fortunately, we can exploit consistency checks to improve our ability to
discriminate spurious glitches from true gravitational-wave events. Requiring
multiple-detector coincidence (Sec.~\ref{ss:coinc}) is one powerful check, but
here we discuss a separate check on waveform consistency for a single
detector's matched-filter output. This waveform consistency check determines
how similar the SNR time series of the data is to the SNR time series expected
from a real signal.

Under the assumption that the signal in the data exactly matches the
matched-filter template up to a constant, it is possible to predict the local
matched-filter \ac{SNR} by computing the template autocorrelation function and
scaling it to the known \ac{SNR}. However, the known \ac{SNR} is a result of
the matched-filter response from two identical but out of phase templates, thus
instead of scaling the autocorrelation function to the \ac{SNR}, a complex
\ac{SNR} series is constructed from the two matched-filter outputs,
 \begin{align}
	z_j(t) &= x_{2j}(t) + i x_{2j+1}(t), \, j \in [0, \NT - 1].
\end{align} 
These are compared to the complex autocorrelation function,
 \begin{align}
	R_j(t) = \int_{-\infty}^{\infty} \mathrm{d} f \frac{|\widetilde{h}_{2j}(f)|^2+|\widetilde{h}_{2j+1}(f)|^2}{S_n(|f|)} e^{2\pi i f t},
\end{align} 
where $t=0$ is chosen to be the peak time, $t_p$. By convention, each
\emph{real} template is normalized such that its autocorrelation is
$\frac{1}{2}$ at the peak time, thus $R_j(0) = 1$.  We compute a signal
consistency test value, $\xi^2$, as a function of time given the complex
\ac{SNR} time series $z_j(t)$, a trigger's peak complex \ac{SNR} $z_j(0)$, and
the autocorrelation function time series $R_j(t)$ as
 \begin{align}\label{eq:autochisq}
	\xi^2_j(t) = |z_j(t) - z_j(0)R_j(t)|^2.
\end{align} 
If the gravitational wave strain data contain only noise (i.e., $\tilde d(f) =
\tilde n(f)$), then (see Appendix~\ref{app:avgderivation} for derivation)
 \begin{align}\label{eq:autochisqvar}
\langle \xi^2_j(t) \rangle = 2 - 2|R_j(t)|^2.
\end{align} 

In practice, a value of $\xi^2$ is computed for each trigger by
integrating $\xi^2(t)$ in a window of time around the trigger and normalizing
it using Eq.~\eqref{eq:autochisqvar}. The integral takes the form
 \begin{align}
\xi_j^2 = \frac{\int_{-\delta t}^{\delta t} \mathrm{d} t |z_j(t) - z_j(0) R_j(t)|^2}{\int_{-\delta t}^{\delta t} \mathrm{d} t (2 - 2|R_j(t)|^2)},
\label{eq:autochisq_integrated}
\end{align} 
where $\delta t$ is a tunable parameter that defines the size of the window
around the peak time over which to perform the integration. 
\begin{figure}[h!!]
\includegraphics[width=0.90\columnwidth]{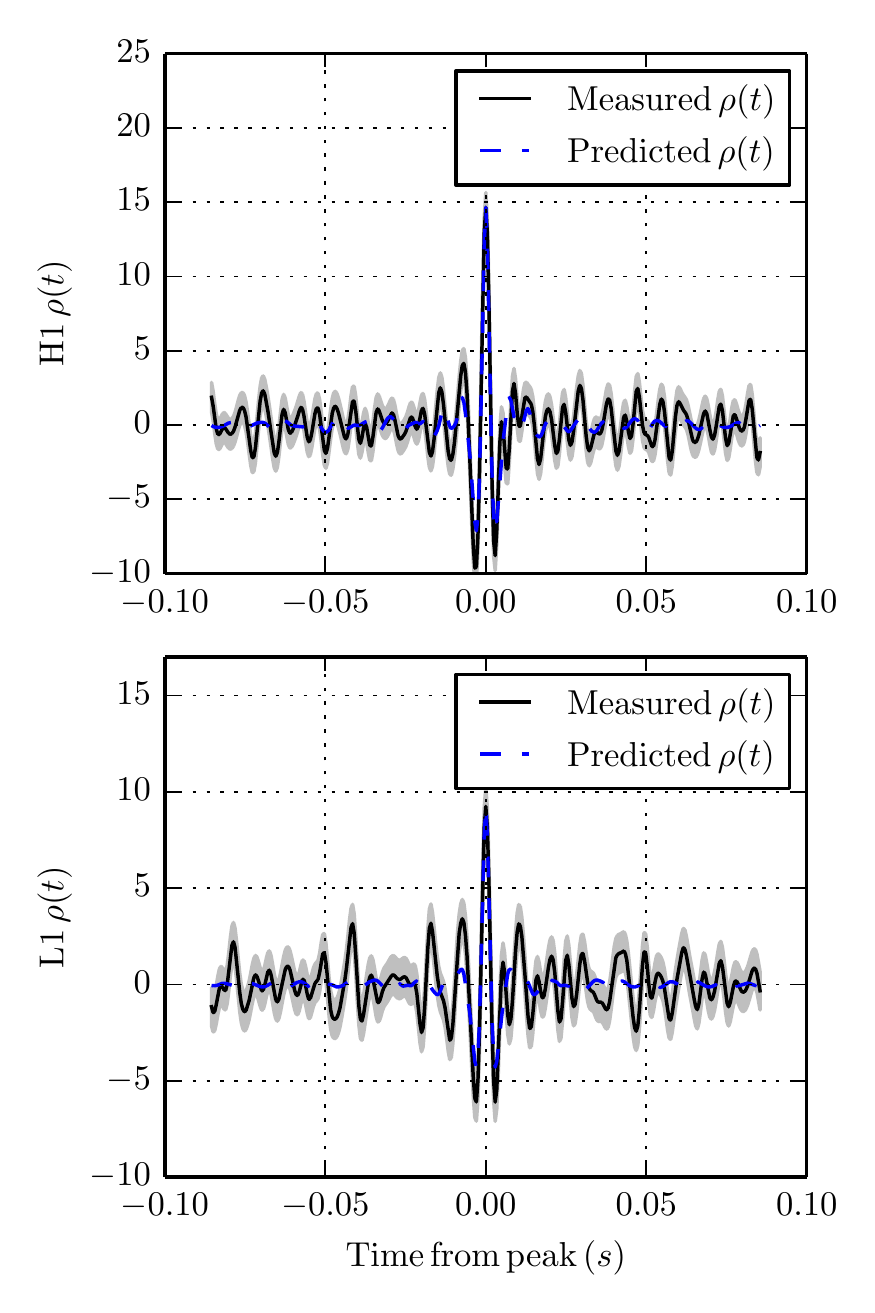}
\caption{Ingredients in the auto-correlation-based least-squares test as
described in \eqref{eq:autochisq}.  The two panels show the \ac{SNR} time
series near a simulated signal in Initial LIGO data (black) along with the
predicted \ac{SNR} computed from the template autocorrelation.  Subtracting
these two time series and integrating their squared magnitude provides a signal
consistency test, $\xi^2$, at the time of a given trigger that can be
used to reject non-stationary noise transients.
\label{f:autoveto}
}
\end{figure}
Typically, $\delta
t$ is calculated in terms of an odd-valued autocorrelation length (ACL),
specified as a number of samples such that $\delta t = (\text{ACL}-1) \Delta
t_s / 2$, where $\Delta t_s = f_s^{-1}$ is the sampling time step. A suitable
value for ACL was found to be 351 samples when filtering is conducted at a
$\unit{2048}{\hertz}$ sample rate, resulting in $\delta t \sim
\unit{85.4}{\milli\second}$; this value was found using by Monte Carlo
simulations in real data.

Fig.~\ref{f:autoveto} plots the \ac{SNR} and scaled autocorrelation for a
template that recovered a simulated signal in Initial LIGO data.  Subtracting
the measured \ac{SNR} time series from the predicted series shown in this
figure is what is done in \eqref{eq:autochisq_integrated} on a
trigger-by-trigger basis.

We note that $\xi^2$ differs from the traditional time-frequency $\chi^2$ test
in~\cite{allen2005chi}, and it is not in fact a $\chi^2$-distributed number in
Gaussian noise.  However, the statistics of the $\xi^2$ test are recorded for
both noise and simulated signals and can therefore be used in the
likelihood-ratio test described in Sec.~\ref{ss:rank}.

%% file: coincidence.tex
Demanding that two triggers are found in temporal coincidence between the LIGO
sites is a powerful technique to suppress the background of the search.  For a
single detector trigger, we define the time of an event to coincide with the
peak of its \ac{SNR} time series. Given a trigger in one detector, we check for
corresponding triggers in the other detector within an appropriate time window,
which takes into account the maximum gravitational-wave travel time between
detectors and statistical fluctuations in the measured event time due to
detector noise.  For the two LIGO detectors, the time window is typically $\pm
15$ ms.  We further require that the mass and spin template parameters are the
same for the two triggers. This exact match requirement potentially results in
a small loss of \ac{SNR} for real signals, since the loudest trigger in each
detector will in general not have the exact same template parameters due to
independent noise in the detectors. However, taking into account such
fluctuations requires detailed knowledge of the metric on the signal
manifold~\cite{robinson2008geometric}, which may not be easily available.
Furthermore, the exact match restriction suppresses the noise and drastically
simplifies the pipeline.

%% file: likelihood.tex
Each trigger from each detector has independently computed $\rho$, $\xi^2$, and
$t_p$ values.  After coincidences are formed, it is necessary to rank the
coincident events from least likely to be a signal to most likely to be a
signal and to assign a significance to each. The \gstlal{}-based inspiral
pipeline uses the likelihood-ratio statistic described in
\cite{cannon2015likelihood} to rank coincident events by their \ac{SNR},
$\xi^2$, the instantaneous sensitivity of each detector (expressed as the
horizon distance, $\horizons$), and the detectors involved in the coincidence
(expressed as the set $\detectors$). For the case where only the \ac{aLIGO}
observatories H1 and L1 are participating, the likelihood ratio of an event
found in coincidence is defined as
\begin{widetext}
 \begin{align}
        \likehood \left( \horizons, \detectors, \snr{H},\chisq{H},\snr{L},\chisq{L},\params \right) &= \likehood \left( \horizons, \detectors, \snr{H},\chisq{H},\snr{L},\chisq{L}  \mid  \params \right) \likehood \left( \params \right) \notag \\
        &= \frac{P \left( \horizons, \detectors, \snr{H},\chisq{H},\snr{L},\chisq{L} \mid  \params, \sh \right)}{P \left( \horizons, \detectors, \snr{H},\chisq{H},\snr{L},\chisq{L} \mid  \params, \nh \right)} \likehood \left( \params \right) \label{eq: likelihood definition},
\end{align} 
\end{widetext}
where $\params$ is a label corresponding to the template bank bin being
matched-filtered (Sec.~\ref{ss:bank}).  The numerator and denominator are
factored into products of several terms in \cite{cannon2015likelihood},
assuming that the noise distributions for each interferometer are independent
of each other. The computation of each term in the factored numerator and
denominator is discussed in detail in \cite{cannon2015likelihood}; in this
paper, we will give only a short summary of the denominator. The denominator is
factored such that 
 \begin{align}
	P &\left( \horizons, \detectors, \snr{H},\chisq{H},\snr{L},\chisq{L} \mid  \params, \nh \right) \notag \\
	&\propto \prod_{\rm{inst} \in \detectors} \cprob{\rho_{\rm{inst}}, \xi^2_{\rm{inst}}}{\params, \nh}.
\end{align}  
The detection statistics $\rho$ and $\xi^2$ from non-coincident triggers are
used to populate histograms for each detector, which are then normalized and
smoothed by a Gaussian smoothing kernel to approximate
$\cprob{\rho_{\rm{inst}}, \xi^2_{\rm{inst}}}{\params, \nh}$. Running in the
low-latency operation mode requires a burn-in period until the analysis
collects enough non-coincident triggers to construct an accurate estimate of
the $(\rho, \xi^2)$ PDFs. Neither operation mode tracks time dependence of
these PDFs, instead the PDFs are constructed from cumulative histograms. Future
work may add time dependence.

Rather than collecting non-coincident $(\rho, \xi^2)$ statistics from
individual templates, we group linearly dependent templates together to avoid
the computational cost and complexity of tracking each template separately.
Furthermore, it has been observed that groups of linearly dependent templates
produce similar PDFs, thus coarse graining the parameter space allows one to
approximate these PDFs for collections of templates. Therefore, the $\params$
in the likelihood ratio is a label that identifies a specific template bank
bin. Exactly how templates are grouped together into background bins is left as
a tuning decision for the user, but typically $\mathcal{O}(1000)$ templates
from each detector are grouped together.

\begin{figure}[h!]
\includegraphics[width=\columnwidth]{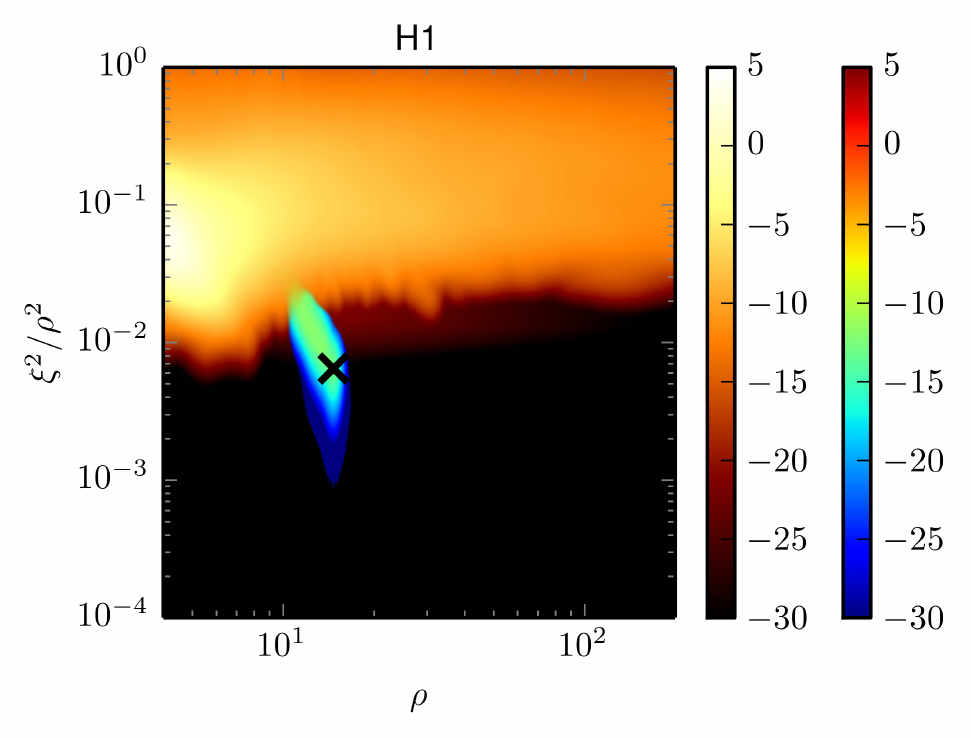}
\includegraphics[width=\columnwidth]{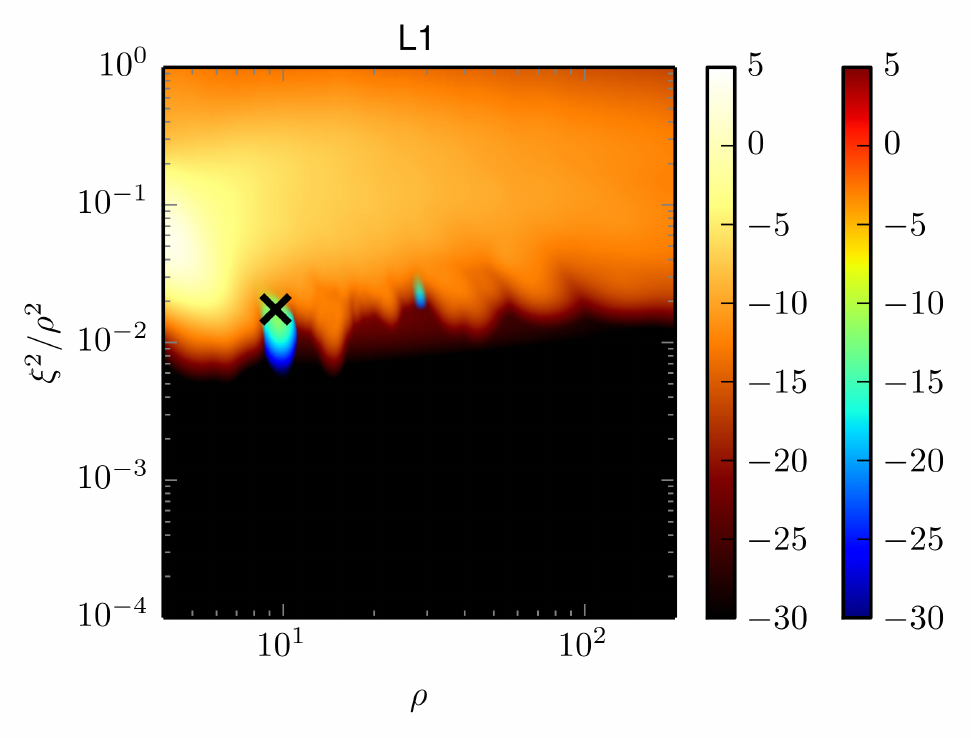}
\caption{\label{f:bigdog_snrchisq} PDFs used in the likelihood ratio
calculation, generated by histogramming, then smoothing, and normalizing the
triggers.  The plots shown are from an analysis of S6 data beginning at
September 14, 2010, at 23:58:48 UTC and ending at September 21, 2010, at
23:58:48 UTC, which includes the blind injection known as the ``Big Dog.'' The
warm colormap corresponds to the natural logarithm of marginalized probability
density function estimated from non-coincident triggers only; the cool-colormap
region was computed by adding coincident triggers to the histograms before
smoothing and normalizing. Regions of the cool colormap model consistent with
the warm-colormap model were then masked.  The location of the Big Dog in the
$(\rho, \xi^2)$ plane is marked with a black X.}
\end{figure}

Examples of the $\rho$ and $\xi^2$ distributions, estimated from an analysis of
one week of S6 data, are shown in Fig.~\ref{f:bigdog_snrchisq}. The analysis
considered data recorded between September 14, 2010, 23:58:48 UTC and September
21, 2010, 23:58:48 UTC.  These boundaries were chosen to include the blind
injection performed on September 16, 2010, at 06:42:23 UTC, often referred to
as the ``Big Dog.'' The warm colormap corresponds to the natural logarithm of
the estimated noise probability density function.  The cool colormap
corresponds to a PDF generated by adding the coincident triggers to the single
detector triggers before smoothing and normalizing. The cool-colormap
distribution was then masked to only show regions which deviate from the
background estimate. The location of the Big Dog parameters is marked with a
black X.

\begin{figure}[h!]
\includegraphics[width=\columnwidth]{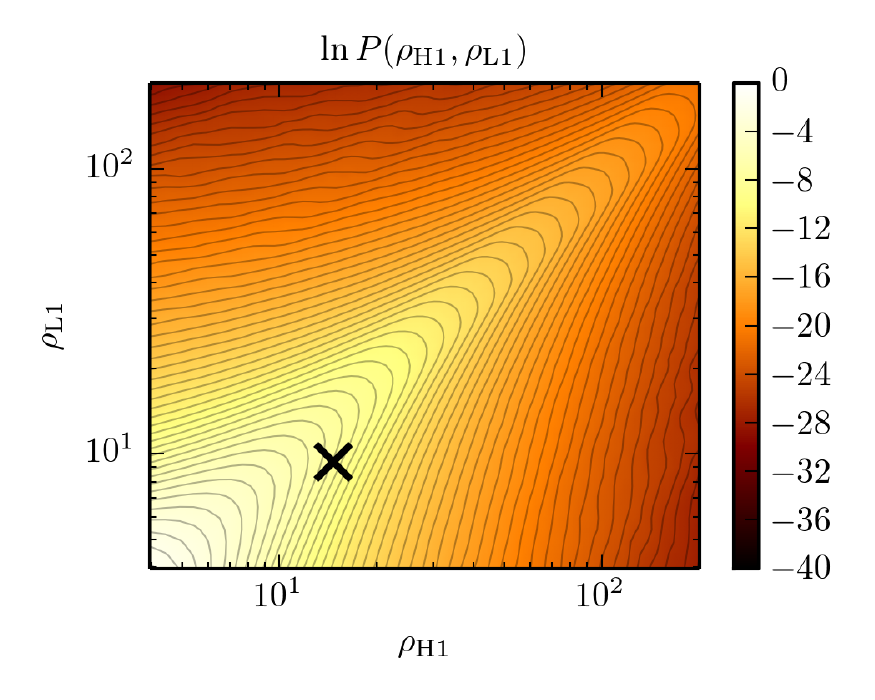}
\includegraphics[width=\columnwidth]{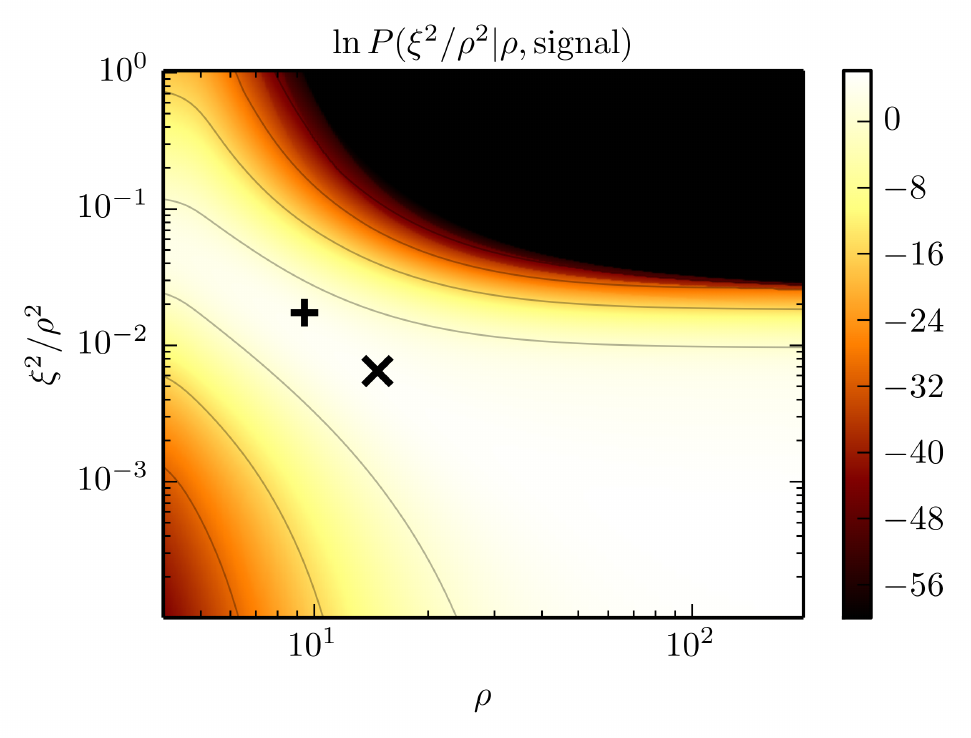}
\caption{\label{f:likelihood_numerator} Instances of two of the distributions
included in the calculation of the likelihood-ratio numerator, generated from
an analysis of S6 data beginning at September 14, 2010, at 23:58:48 UTC and
ending at September 21, 2010, at 23:58:48 UTC. Top: The joint SNR PDF used to
enforce amplitude consistency across observatories. The location of the
measured Big Dog parameters is marked with a black X. Bottom: The $(\rho,
\xi^2)$ signal distribution used in the numerator of the likelihood ratio. The
locations of the measured Big Dog parameters are marked with a black X for
Hanford and a black + for Livingston.}
\end{figure}

Examples of two other likelihood-ratio components from the Big Dog analysis are
shown in Fig.~\ref{f:likelihood_numerator}. The top plot shows the joint
\ac{SNR} PDF, which is used in the numerator of the likelihood ratio to enforce
amplitude consistency~\cite{cannon2015likelihood}; the bottom plot shows the
signal hypothesis model of the $(\rho, \xi^2)$ plane. The semi-analytic models
used to generate these plots are described in \cite{cannon2015likelihood}.

%% file: datareduction.tex
Signals can produce several high-likelihood events at the same time in
different templates; we wish to ensure that we only consider the most likely
event associated with a signal.  In the offline analysis, we use a clustering
algorithm that picks out the maximum likelihood-ratio event globally across the
input template bank within a $\pm 4$-second window.  The online analysis does
not cluster events globally to reduce latency. Instead, the online analysis
keeps the maximum likelihood-ratio event in each $\params$ bin within a $\pm
1$-second window.

%% file: eventprocessing.tex
The result of the pipeline components described in Sec.~\ref{sec:ident} is a
list of events ranked from most to least likely to be a gravitational-wave
signal. In this section, we discuss how the significance is estimated, the
procedure in the case that a sufficiently significant candidate is identified,
and how simulated waveforms are used to to characterize the sensitivity of the
analysis to gravitational waves.

%% file: significance.tex
Most coincident events are noise, thus the p-value, the probability that noise
would produce an event with a ranking statistic at least as large as the one
under consideration, is the standard tool used to identify candidate
gravitational-wave events. The p-value has conventionally been evaluated by
performing \textit{time slides}, where a set of time-shifts that are much
larger than the gravitational-wave travel time (tens of milliseconds) between
gravitational-wave detectors is introduced into one or more datasets and the
coincidence and event-ranking procedure is repeated in the same way as it is
done without the time-shifts~\cite{capano2011searching}. Instead of performing
time slides, the \gstlal{}-based inspiral pipeline uses triggers not found in
coincidence to compute a kernel density estimate of the probability density of
noise-like events in each background bin, $P \left( \ln \likehood \mid \params,
\nh \right)$ \cite{cannon2013method, cannon2015likelihood}. The background bins
are then marginalized over to obtain $P \left( \ln \likehood \mid \nh \right)$
and the complementary cumulative distribution,
 \begin{align}
	C \left( \ln \likehood^* \mid \nh \right) &= \int_{\ln \likehood^*}^\infty \mathrm{d} \ln \likehood \; P \left( \ln \likehood \mid \nh \right).\label{eq: ccdf}
\end{align} 

The p-value we seek describes the probability that a population of $M$
independent coincident noise-like events contains at least one event with a log
likelihood ratio greater than or equal to some threshold $\ln \likehood^*$.
This can be written as the complement of the binomial
distribution~\cite{cannon2013method},
 \begin{align}
	P \left( \ln \likehood \right. &\ge \left.\ln \likehood^* \mid \nh_1, \dots, \nh_M \right)  \notag \\
	&= 1 - \binom{M}{0} \left(1 - e^{-C \left( \ln \likehood^* \mid \nh \right)} \right)^0 \times \notag \\
	&\left(e^{-C \left( \ln \likehood^* \mid \nh \right)} \right)^M \notag \\
	&= 1 - e^{-M C \left( \ln \likehood^* \mid \nh \right)}, \label{eq: FAP}
\end{align} 
where $e^{-C \left( \ln \likehood^* \mid \nh \right)}$ is the probability that
a Poisson process with mean rate $C \left( \ln \likehood^* \mid \nh \right)$
will yield an event with log likelihood ratio less than $\ln \likehood^*$. The
binomial coefficient and the term that follows are both clearly unity and were
only explicitly written for pedagogical reasons.

When calculating an event's significance during an experiment of undetermined
length, such as the low-latency processing of data during a science run, it is
convenient to express the significance in terms of how often the noise is
expected to yield an event with a log likelihood ratio $\ge \ln \likehood^*$.
This is referred to as the \ac{FAR}~\cite{cannon2013method}; for an experiment
of length $T$, we define this as
 \begin{align} \rm{FAR} &= \frac{ C
	\left( \ln \likehood^* \mid \nh \right)}{T} .\label{eq: FAR}
\end{align}  
The time used in the calculation of the \ac{FAR} is the total elapsed observing
time regardless of instrument state for the low-latency configuration and the
total time where at least two detectors are operating for the offline analysis
operation.  The offline configuration definition is historically what has been
used; however, the online definition leads to intuitive false alarm rates for
sharing low-latency events with external observing partners.

The procedure to estimate the background distribution described thus far does not
account for the clustering described in Sec.~\ref{ss:cluster}. Events with low
$\ln \likehood$ are more common than those with high $\ln \likehood$, thus the
clustering process removes low $\ln \likehood$ events preferentially. The
normalization of the background model is determined by the observed events
above a log-likelihood ratio threshold chosen to be safely out of the region
affected by clustering. This is acceptable because low $\ln \likehood$ events
are, by construction, the least likely to contain a signal. Consequently,
we only consider events well above this threshold as viable candidates. Work is
currently underway to create a background model that accounts for clustering.

\begin{figure}[!h]
\includegraphics[width=\columnwidth]{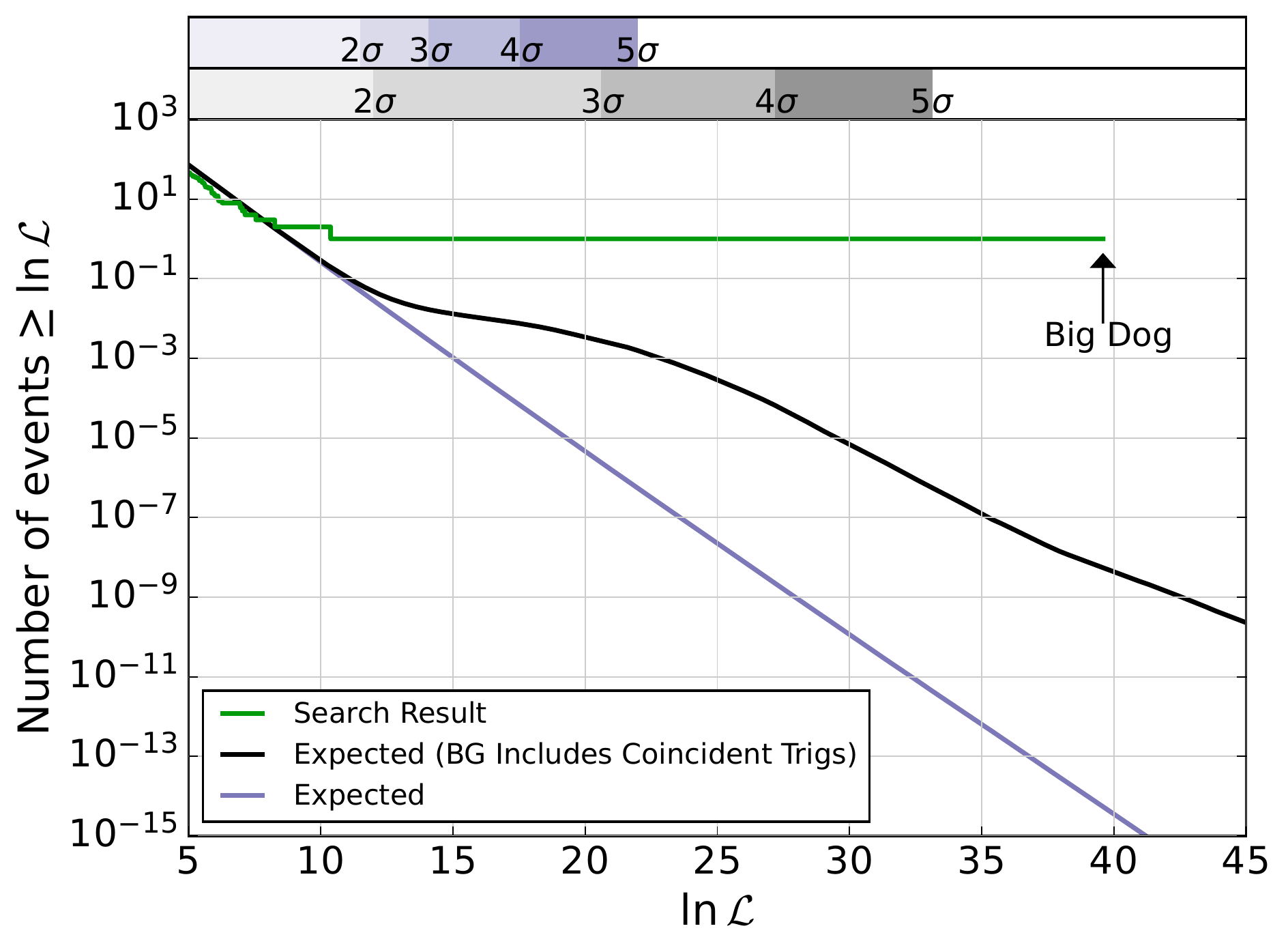}
\caption{\label{f:bigdog_result} Number of observed events as a function of log
likelihood ratio in an analysis of S6 data beginning on September 14, 2010, at
23:58:48 UTC and ending at September 21, 2010, at 23:58:48 UTC. The Big Dog
injection, found with a false alarm probability of \bigdogfap{} (\bigdogsigma),
is marked on the observed distribution (green). The black line represents the
predicted number of events when observed events are included in the background
model, while the blue line is the predicted number when the observed events are not
included in the background model.}
\end{figure}

A plot of the significance results from the Big Dog run discussed in
Sec.~\ref{ss:rank} is shown in Fig.~\ref{f:bigdog_result}. The Big Dog was
found with a p-value of \bigdogfap{} (\bigdogsigma), which corresponded to a FAR of
\bigdogfar~Hz (1 per $\sim \bigdogifar$~years); Table~\ref{t:bigdog_params}
lists the recovered parameters of the Big Dog.

\input{bigdogtable.tex}

%% file: bigdogtable.tex
\begin{table}
\begin{tabular}{c | c}
p-value   & \bigdogfap{} (\bigdogsigma) \\
\hline
FAR (Hz) & \bigdogfar \\
\hline 
$\log \mathcal{L}$  & \bigdoglikelihood \\
\hline 
$\rho_{\text{H}}$ & \bigdoghsnr \\
\hline
$\xi^2_{\text{H}}$  & \bigdoghxisq \\
\hline
$\rho_{\text{L}}$ & \bigdoglsnr \\
\hline 
$\xi^2_{\text{L}}$  & \bigdoglxisq \\
\hline
\mchirp (M$_\odot$) & \bigdogmchirp \\
\end{tabular}
\caption{\label{t:bigdog_params} The result of the analysis of S6 data
beginning on September 14, 2010, at 23:58:48 UTC and ending on September 21,
2010, at 23:58:48 UTC. Only the parameters found for the recovered Big Dog
injection are shown. The Big Dog was the most significant event found in this
analysis period.}
\end{table}

%% file: alerts.tex
When operating in a low-latency analysis configuration, one of the primary
goals of the \gstlal{}-based inspiral pipeline is to identify candidate events
and upload them to the Gravitational-wave Candidate Event Database
(\gracedb{}~\cite{gracedb}) as quickly as possible in order to issue alerts to
observing partners~\cite{GWEMAlerts}.  

Events that pass a given \ac{FAR} threshold are identified within $\sim 1$
minute of the gravitational-wave signals arriving at Earth.  The basic
parameters of the event are transmitted to \gracedb{}, including the GPS time
of the event, the \ac{SNR} and $\xi^2$ values for the triggers in each detector, and
the parameters of the best-fit template (for example, mass and spin values, the
significance estimate, etc.).  Furthermore, the instantaneous estimate of the
\ac{PSD} is uploaded to \gracedb{}, as well as the histogram data used in
computing the p-value.

An event upload automatically initiates several automated and human followup
activities to aid rapid communication with observing partners~\cite{lvalert}.
First, a rapid sky localization routine known as
BAYESTAR~\cite{singer2014first,berry2014parameter,singer2016rapid} uses the
event information and the \ac{PSD} to estimate the event's sky position within
minutes.  At the same time, deeper parameter estimation analysis begins in
order to provide updated position reconstruction, as well as the full posterior
probability distributions of the binary parameters~\cite{veitch2015parameter},
on a timescale that ranges from hours to days.

In addition to parameter estimation, data-quality information is also mined to
provide rapid feedback to analysts.  Time-frequency spectrograms are
automatically generated to indicate the stationarity of noise near an
event~\cite{ligodv}.  Furthermore, low-latency mining of LIGO's auxiliary
channels provide additional information about the state of the detector and
environment when an alert is first
generated~\cite{biswas2013application,essick2013optimizing,lalsuite}

The suitability of the low-latency pipeline for generating data for external
alerts has been studied extensively in~\cite{singer2014first,
berry2014parameter}

%% file: injections.tex
Simulated gravitational waveforms known as ``software injections'' are used to
assess the pipeline response to real gravitational-wave signals. The LIGO
strain data is duplicated and simulated compact binary waveforms are digitally
added to the duplicated data streams. In low latency, the new data with
software injections added are broadcast to the LIGO Data Grid in parallel to
the normal dataset so that a simultaneous run can measure the instantaneous
sensitivity of the low-latency analysis to compact binary sources.  In the
offline mode, strain data is read from disk, software injections are added, and
the new data is written back to disk before the offline inspiral pipeline
processes the data.

Injections are considered `found' if a coincident event with the correct
template parameters is found with a \ac{FAR}~$\le \unit{30}{\dday}$ at the time
of the injection, and `missed' otherwise. The volume of space the pipeline is
sensitive to, $V$, is approximated as a sphere and computed via
 \begin{align}
	V &= 4 \pi \int_0^\infty \mathrm{d} r \epsilon(r) r^2,
\end{align} 
where $\epsilon(r)$ is an efficiency parameter given by the ratio of found to
total injections modeled to be a distance $r$ away.  We define our estimated
range, the average furthest distance a signal can originate from and still be
detected, as
 \begin{align}
	R &= \left(\frac{3V}{4\pi}\right)^{1/3}. \label{eq:range}
\end{align} 
It is important to note the range depends on parameters of the compact binary
system. For example, the range for a $1.4-1.4 \msun$ binary neutron star system
will be different than that of a $10-10 \msun$ binary black hole system, thus
different injection sets must be used to determine the pipeline's sensitivity
to different regions of the compact binary parameter space.
Eq.~\eqref{eq:range} is compared to the analytically computed SenseMon range,
an estimate of pipeline sensitivity calculated from the
\ac{PSD}~\cite{abadie2010sensitivity}. Comparing the sensitivity estimated from
the \ac{PSD} to the sensitivity estimated from injections provides additional
confidence in the sensitivity estimates.

Typically, injections are added at a much higher rate than the expected
gravitational wave signal rate.  However, their cadence is chosen such that
they do not bias the \ac{PSD} estimate described in Sec.~\ref{ss:psd}. In
practice, injections are typically added about once per minute so that it is
possible to evaluate the average response to certain signal types over the
entire experiment duration.

%% file: conclusion.tex
The \gstlal{}-based inspiral pipeline is a stream-based pipeline that allows
for time-domain compact binary searches capable of identifying and uploading
candidate gravitational-wave signals within seconds. This provides rapid
feedback to the gravitational-wave detector control rooms and enables prompt
event alerts for electromagnetic followup by observing partners.  The analysis
techniques were designed for second- and third-generation gravitational-wave
detectors and have been demonstrated to be applicable even to the
computationally challenging case of the future Einstein
Telescope~\cite{meacher2016second}.

\gstlal{} and all related software is available for public use and
licensed under the GPL~\cite{gstlal}.

%% file: acknowledgements.tex
The authors wish to thank B. Sathyaprakash, the LIGO Scientific Collaboration,
and the Compact Binary Coalescence working group for many useful discussions.

We gratefully acknowledge the support of the Eberly Research Funds of Penn
State and the National Science Foundation through PHY-0757058, NSF-0923409,
PHY-1104371, PHY-1454389, and PHY-1307429.  This document has LIGO document
number: P1600009.

%% file: appendix1.tex
\section{Expectation Value of Signal Consistency Test Value in Noise} \label{app:avgderivation}
Expanding Eq.~\eqref{eq:autochisq} and taking the ensemble average, we find
 \begin{align}
	\langle \xi^2_j (t) \rangle &= \langle |z_j(t) - z_j(0)R_j(t)|^2 \rangle, \notag \\
	&= \langle |z_j(t)|^2 \rangle - 2 \mathrm{Re} \left[ \langle z_j^*(t) z_j(0) \rangle R_j(t) \right] \notag \\
	& + \langle |z_j(0)|^2 \rangle |R_j(t)|^2.
\end{align} 
Starting with Eq.~\eqref{eq:gstlalfreqmf}, 
 \begin{align} 
  \langle | z_j(t) |^2 \rangle &= \left\langle \left| 2 \int_{-\infty}^\infty \mathrm{d}f \frac{ \widetilde{n}(f) ( \widetilde{h}_{2j}^*(f) + i \widetilde{h}_{2j+1}^*(f)) }{S_n(f)} e ^{ 2 \pi i t f }  \right|^2 \right\rangle, \notag \\
	&= 4 \int_{-\infty}^\infty \int_{-\infty}^\infty \mathrm{d} f_1 \mathrm{d}f_2 \left( \left\langle \widetilde{n}(f_1) \widetilde{n}(f_2) \right\rangle e ^{ 2 \pi i t (f_1 - f_2) } \vphantom{\frac{\widetilde{h}_{2j}^*(f_1)}{S_n(|f_1|)}} \right. \notag \\
	& \left. \frac{ (\widetilde{h}_{2j}^*(f_1) +i \widetilde{h}_{2j+1}^*(f)) ( \widetilde{h}_{2j} (f_2) - i \widetilde{h}_{2j+1} (f_2))}{S_n(|f_1|) S_n(|f_2|) } \right), \notag \\
	&= 2 \int_{-\infty}^\infty \mathrm{d} f \frac{ | \widetilde{h}_{2j}(f)|^2 + |\widetilde{h}_{2j+1}(f)|^2}{S_n(|f|) }, \notag \\
	\langle | z_j(t) |^2 \rangle &= \langle | z_j(0) |^2 \rangle = 2.
\end{align} 
where Eq.~\eqref{eq:sspsd} was used in the last step. Computing $\langle
z_j^*(t) z_j(0) \rangle$ follows the same steps, except the $z_j(0)$ term does
have a complex exponential to cancel the complex exponential accompanying
$z_j^*(t)$, thus 
 \begin{align}
	\langle z_j^*(t) z_j(0) \rangle &= 2 R_j^*(t), \\
	\langle \xi^2_j (t) \rangle &= 2 - 2 |R_j(t)|^2
\end{align}

%% file: methods.bbl
%

%% file: acronyms.tex
\begin{acronym}
\acro{GW}{gravitational-wave}
\acro{LSC}{LIGO Scientific Collaboration}
\acro{PSD}{power spectral density}
\acro{SNR}{signal-to-noise ratio}
\acro{SVD}{singular value decomposition}
\acro{aLIGO}{advanced LIGO}
\acro{FAR}{false-alarm rate}
\acro{FAP}{false-alarm probability}
\end{acronym}

%% file: methods.bbl
\begin{thebibliography}{60}%
\makeatletter
\providecommand \@ifxundefined [1]{%
 \@ifx{#1\undefined}
}%
\providecommand \@ifnum [1]{%
 \ifnum #1\expandafter \@firstoftwo
 \else \expandafter \@secondoftwo
 \fi
}%
\providecommand \@ifx [1]{%
 \ifx #1\expandafter \@firstoftwo
 \else \expandafter \@secondoftwo
 \fi
}%
\providecommand \natexlab [1]{#1}%
\providecommand \enquote  [1]{``#1''}%
\providecommand \bibnamefont  [1]{#1}%
\providecommand \bibfnamefont [1]{#1}%
\providecommand \citenamefont [1]{#1}%
\providecommand \href@noop [0]{\@secondoftwo}%
\providecommand \href [0]{\begingroup \@sanitize@url \@href}%
\providecommand \@href[1]{\@@startlink{#1}\@@href}%
\providecommand \@@href[1]{\endgroup#1\@@endlink}%
\providecommand \@sanitize@url [0]{\catcode `\\12\catcode `\$12\catcode
  `\&12\catcode `\#12\catcode `\^12\catcode `\_12\catcode `\%12\relax}%
\providecommand \@@startlink[1]{}%
\providecommand \@@endlink[0]{}%
\providecommand \url  [0]{\begingroup\@sanitize@url \@url }%
\providecommand \@url [1]{\endgroup\@href {#1}{\urlprefix }}%
\providecommand \urlprefix  [0]{URL }%
\providecommand \Eprint [0]{\href }%
\providecommand \doibase [0]{http://dx.doi.org/}%
\providecommand \selectlanguage [0]{\@gobble}%
\providecommand \bibinfo  [0]{\@secondoftwo}%
\providecommand \bibfield  [0]{\@secondoftwo}%
\providecommand \translation [1]{[#1]}%
\providecommand \BibitemOpen [0]{}%
\providecommand \bibitemStop [0]{}%
\providecommand \bibitemNoStop [0]{.\EOS\space}%
\providecommand \EOS [0]{\spacefactor3000\relax}%
\providecommand \BibitemShut  [1]{\csname bibitem#1\endcsname}%
\let\auto@bib@innerbib\@empty
\bibitem [{\citenamefont {Abbott}\ \emph
  {et~al.}(2016{\natexlab{a}})\citenamefont {Abbott}, \citenamefont {Abbott},
  \citenamefont {Abbott}, \citenamefont {Abernathy}, \citenamefont {Acernese},
  \citenamefont {Ackley}, \citenamefont {Adams}, \citenamefont {Adams},
  \citenamefont {Addesso}, \citenamefont {Adhikari} \emph
  {et~al.}}]{abbott2016observation}%
  \BibitemOpen
  \bibfield  {author} {\bibinfo {author} {\bibfnamefont {B.}~\bibnamefont
  {Abbott}}, \bibinfo {author} {\bibfnamefont {R.}~\bibnamefont {Abbott}},
  \bibinfo {author} {\bibfnamefont {T.}~\bibnamefont {Abbott}}, \bibinfo
  {author} {\bibfnamefont {M.}~\bibnamefont {Abernathy}}, \bibinfo {author}
  {\bibfnamefont {F.}~\bibnamefont {Acernese}}, \bibinfo {author}
  {\bibfnamefont {K.}~\bibnamefont {Ackley}}, \bibinfo {author} {\bibfnamefont
  {C.}~\bibnamefont {Adams}}, \bibinfo {author} {\bibfnamefont
  {T.}~\bibnamefont {Adams}}, \bibinfo {author} {\bibfnamefont
  {P.}~\bibnamefont {Addesso}}, \bibinfo {author} {\bibfnamefont
  {R.}~\bibnamefont {Adhikari}},  \emph {et~al.},\ }\href@noop {} {\bibfield
  {journal} {\bibinfo  {journal} {Physical Review Letters}\ }\textbf {\bibinfo
  {volume} {116}},\ \bibinfo {pages} {061102} (\bibinfo {year}
  {2016}{\natexlab{a}})}\BibitemShut {NoStop}%
\bibitem [{\citenamefont {Abbott}\ \emph
  {et~al.}(2016{\natexlab{b}})\citenamefont {Abbott}, \citenamefont {Abbott},
  \citenamefont {Abbott}, \citenamefont {Abernathy}, \citenamefont {Acernese},
  \citenamefont {Ackley}, \citenamefont {Adams}, \citenamefont {Adams},
  \citenamefont {Addesso}, \citenamefont {Adhikari} \emph
  {et~al.}}]{abbott2016gw151226}%
  \BibitemOpen
  \bibfield  {author} {\bibinfo {author} {\bibfnamefont {B.}~\bibnamefont
  {Abbott}}, \bibinfo {author} {\bibfnamefont {R.}~\bibnamefont {Abbott}},
  \bibinfo {author} {\bibfnamefont {T.}~\bibnamefont {Abbott}}, \bibinfo
  {author} {\bibfnamefont {M.}~\bibnamefont {Abernathy}}, \bibinfo {author}
  {\bibfnamefont {F.}~\bibnamefont {Acernese}}, \bibinfo {author}
  {\bibfnamefont {K.}~\bibnamefont {Ackley}}, \bibinfo {author} {\bibfnamefont
  {C.}~\bibnamefont {Adams}}, \bibinfo {author} {\bibfnamefont
  {T.}~\bibnamefont {Adams}}, \bibinfo {author} {\bibfnamefont
  {P.}~\bibnamefont {Addesso}}, \bibinfo {author} {\bibfnamefont
  {R.}~\bibnamefont {Adhikari}},  \emph {et~al.},\ }\href@noop {} {\bibfield
  {journal} {\bibinfo  {journal} {Physical Review Letters}\ }\textbf {\bibinfo
  {volume} {116}},\ \bibinfo {pages} {241103} (\bibinfo {year}
  {2016}{\natexlab{b}})}\BibitemShut {NoStop}%
\bibitem [{\citenamefont {Aasi}\ \emph {et~al.}(2015)\citenamefont {Aasi},
  \citenamefont {Abbott}, \citenamefont {Abbott}, \citenamefont {Abbott},
  \citenamefont {Abernathy}, \citenamefont {Ackley}, \citenamefont {Adams},
  \citenamefont {Adams}, \citenamefont {Addesso}, \citenamefont {Adhikari}
  \emph {et~al.}}]{aasi2015advanced}%
  \BibitemOpen
  \bibfield  {author} {\bibinfo {author} {\bibfnamefont {J.}~\bibnamefont
  {Aasi}}, \bibinfo {author} {\bibfnamefont {B.}~\bibnamefont {Abbott}},
  \bibinfo {author} {\bibfnamefont {R.}~\bibnamefont {Abbott}}, \bibinfo
  {author} {\bibfnamefont {T.}~\bibnamefont {Abbott}}, \bibinfo {author}
  {\bibfnamefont {M.}~\bibnamefont {Abernathy}}, \bibinfo {author}
  {\bibfnamefont {K.}~\bibnamefont {Ackley}}, \bibinfo {author} {\bibfnamefont
  {C.}~\bibnamefont {Adams}}, \bibinfo {author} {\bibfnamefont
  {T.}~\bibnamefont {Adams}}, \bibinfo {author} {\bibfnamefont
  {P.}~\bibnamefont {Addesso}}, \bibinfo {author} {\bibfnamefont
  {R.}~\bibnamefont {Adhikari}},  \emph {et~al.},\ }\href@noop {} {\bibfield
  {journal} {\bibinfo  {journal} {Classical and quantum gravity}\ }\textbf
  {\bibinfo {volume} {32}},\ \bibinfo {pages} {074001} (\bibinfo {year}
  {2015})}\BibitemShut {NoStop}%
\bibitem [{\citenamefont {Acernese}\ \emph {et~al.}(2015)\citenamefont
  {Acernese}, \citenamefont {Agathos}, \citenamefont {Agatsuma}, \citenamefont
  {Aisa}, \citenamefont {Allemandou}, \citenamefont {Allocca}, \citenamefont
  {Amarni}, \citenamefont {Astone}, \citenamefont {Balestri}, \citenamefont
  {Ballardin} \emph {et~al.}}]{acernese2015advanced}%
  \BibitemOpen
  \bibfield  {author} {\bibinfo {author} {\bibfnamefont {F.}~\bibnamefont
  {Acernese}}, \bibinfo {author} {\bibfnamefont {M.}~\bibnamefont {Agathos}},
  \bibinfo {author} {\bibfnamefont {K.}~\bibnamefont {Agatsuma}}, \bibinfo
  {author} {\bibfnamefont {D.}~\bibnamefont {Aisa}}, \bibinfo {author}
  {\bibfnamefont {N.}~\bibnamefont {Allemandou}}, \bibinfo {author}
  {\bibfnamefont {A.}~\bibnamefont {Allocca}}, \bibinfo {author} {\bibfnamefont
  {J.}~\bibnamefont {Amarni}}, \bibinfo {author} {\bibfnamefont
  {P.}~\bibnamefont {Astone}}, \bibinfo {author} {\bibfnamefont
  {G.}~\bibnamefont {Balestri}}, \bibinfo {author} {\bibfnamefont
  {G.}~\bibnamefont {Ballardin}},  \emph {et~al.},\ }\href@noop {} {\bibfield
  {journal} {\bibinfo  {journal} {Classical and Quantum Gravity}\ }\textbf
  {\bibinfo {volume} {32}},\ \bibinfo {pages} {024001} (\bibinfo {year}
  {2015})}\BibitemShut {NoStop}%
\bibitem [{\citenamefont {Aso}\ \emph {et~al.}(2013)\citenamefont {Aso},
  \citenamefont {Michimura}, \citenamefont {Somiya}, \citenamefont {Ando},
  \citenamefont {Miyakawa}, \citenamefont {Sekiguchi}, \citenamefont
  {Tatsumi},\ and\ \citenamefont {Yamamoto}}]{aso2013interferometer}%
  \BibitemOpen
  \bibfield  {author} {\bibinfo {author} {\bibfnamefont {Y.}~\bibnamefont
  {Aso}}, \bibinfo {author} {\bibfnamefont {Y.}~\bibnamefont {Michimura}},
  \bibinfo {author} {\bibfnamefont {K.}~\bibnamefont {Somiya}}, \bibinfo
  {author} {\bibfnamefont {M.}~\bibnamefont {Ando}}, \bibinfo {author}
  {\bibfnamefont {O.}~\bibnamefont {Miyakawa}}, \bibinfo {author}
  {\bibfnamefont {T.}~\bibnamefont {Sekiguchi}}, \bibinfo {author}
  {\bibfnamefont {D.}~\bibnamefont {Tatsumi}}, \ and\ \bibinfo {author}
  {\bibfnamefont {H.}~\bibnamefont {Yamamoto}},\ }\href@noop {} {\bibfield
  {journal} {\bibinfo  {journal} {Physical Review D}\ }\textbf {\bibinfo
  {volume} {88}},\ \bibinfo {pages} {043007} (\bibinfo {year}
  {2013})}\BibitemShut {NoStop}%
\bibitem [{\citenamefont {Iyer}\ \emph {et~al.}(2011)\citenamefont {Iyer} \emph
  {et~al.}}]{LigoIndia}%
  \BibitemOpen
  \bibfield  {author} {\bibinfo {author} {\bibfnamefont {B.}~\bibnamefont
  {Iyer}} \emph {et~al.},\ }\href {{https://dcc.ligo.org/LIGO-M1100296/public}}
  {\enquote {\bibinfo {title} {{LIGO-India, Proposal of the Consortium for
  Indian Initiative in Gravita\ tional-wave Observations (IndIGO)}},}\ }
  (\bibinfo {year} {2011}),\ \bibinfo {note} {{LIGO-DCC-M1100296}}\BibitemShut
  {NoStop}%
\bibitem [{\citenamefont {{LIGO RATES PAPER, TO BE ADDED BEFORE
  SUBMISSION}}()}]{RATESPAPER}%
  \BibitemOpen
  \bibfield  {author} {\bibinfo {author} {\bibnamefont {{LIGO RATES PAPER, TO
  BE ADDED BEFORE SUBMISSION}}},\ }\href@noop {} {\bibinfo  {journal} {In
  review}\ }\BibitemShut {NoStop}%
\bibitem [{\citenamefont {Abadie}\ \emph
  {et~al.}(2010{\natexlab{a}})\citenamefont {Abadie}, \citenamefont {Abbott},
  \citenamefont {Abbott}, \citenamefont {Abernathy}, \citenamefont {Accadia},
  \citenamefont {Acernese}, \citenamefont {Adams}, \citenamefont {Adhikari},
  \citenamefont {Ajith}, \citenamefont {Allen} \emph
  {et~al.}}]{abadie2010predictions}%
  \BibitemOpen
\bibfield  {journal} {  }\bibfield  {author} {\bibinfo {author} {\bibfnamefont
  {J.}~\bibnamefont {Abadie}}, \bibinfo {author} {\bibfnamefont
  {B.}~\bibnamefont {Abbott}}, \bibinfo {author} {\bibfnamefont
  {R.}~\bibnamefont {Abbott}}, \bibinfo {author} {\bibfnamefont
  {M.}~\bibnamefont {Abernathy}}, \bibinfo {author} {\bibfnamefont
  {T.}~\bibnamefont {Accadia}}, \bibinfo {author} {\bibfnamefont
  {F.}~\bibnamefont {Acernese}}, \bibinfo {author} {\bibfnamefont
  {C.}~\bibnamefont {Adams}}, \bibinfo {author} {\bibfnamefont
  {R.}~\bibnamefont {Adhikari}}, \bibinfo {author} {\bibfnamefont
  {P.}~\bibnamefont {Ajith}}, \bibinfo {author} {\bibfnamefont
  {B.}~\bibnamefont {Allen}},  \emph {et~al.},\ }\href@noop {} {\bibfield
  {journal} {\bibinfo  {journal} {Classical and Quantum Gravity}\ }\textbf
  {\bibinfo {volume} {27}},\ \bibinfo {pages} {173001} (\bibinfo {year}
  {2010}{\natexlab{a}})}\BibitemShut {NoStop}%
\bibitem [{\citenamefont {Singer}\ \emph {et~al.}(2014)\citenamefont {Singer},
  \citenamefont {Price}, \citenamefont {Farr}, \citenamefont {Urban},
  \citenamefont {Pankow}, \citenamefont {Vitale}, \citenamefont {Veitch},
  \citenamefont {Farr}, \citenamefont {Hanna}, \citenamefont {Cannon} \emph
  {et~al.}}]{singer2014first}%
  \BibitemOpen
  \bibfield  {author} {\bibinfo {author} {\bibfnamefont {L.~P.}\ \bibnamefont
  {Singer}}, \bibinfo {author} {\bibfnamefont {L.~R.}\ \bibnamefont {Price}},
  \bibinfo {author} {\bibfnamefont {B.}~\bibnamefont {Farr}}, \bibinfo {author}
  {\bibfnamefont {A.~L.}\ \bibnamefont {Urban}}, \bibinfo {author}
  {\bibfnamefont {C.}~\bibnamefont {Pankow}}, \bibinfo {author} {\bibfnamefont
  {S.}~\bibnamefont {Vitale}}, \bibinfo {author} {\bibfnamefont
  {J.}~\bibnamefont {Veitch}}, \bibinfo {author} {\bibfnamefont {W.~M.}\
  \bibnamefont {Farr}}, \bibinfo {author} {\bibfnamefont {C.}~\bibnamefont
  {Hanna}}, \bibinfo {author} {\bibfnamefont {K.}~\bibnamefont {Cannon}},
  \emph {et~al.},\ }\href@noop {} {\bibfield  {journal} {\bibinfo  {journal}
  {The Astrophysical Journal}\ }\textbf {\bibinfo {volume} {795}},\ \bibinfo
  {pages} {105} (\bibinfo {year} {2014})}\BibitemShut {NoStop}%
\bibitem [{\citenamefont {Abbott}\ \emph
  {et~al.}(2016{\natexlab{c}})\citenamefont {Abbott} \emph
  {et~al.}}]{abbott2016emfollow}%
  \BibitemOpen
  \bibfield  {author} {\bibinfo {author} {\bibfnamefont {B.~P.}\ \bibnamefont
  {Abbott}} \emph {et~al.} (\bibinfo {collaboration} {LIGO Scientific,
  Virgo}),\ }\href@noop {} {\  (\bibinfo {year} {2016}{\natexlab{c}})},\
  \Eprint {http://arxiv.org/abs/1602.08492} {arXiv:1602.08492 [astro-ph.HE]}
  \BibitemShut {NoStop}%
\bibitem [{\citenamefont {Adri{\'a}n-Mart{\'{\i}}nez}\ \emph
  {et~al.}(2016)\citenamefont {Adri{\'a}n-Mart{\'{\i}}nez} \emph
  {et~al.}}]{adrian2016neutrino}%
  \BibitemOpen
  \bibfield  {author} {\bibinfo {author} {\bibfnamefont {S.}~\bibnamefont
  {Adri{\'a}n-Mart{\'{\i}}nez}} \emph {et~al.} (\bibinfo {collaboration}
  {ANTARES, IceCube, LIGO Scientific, Virgo}),\ }\href@noop {} {\  (\bibinfo
  {year} {2016})},\ \Eprint {http://arxiv.org/abs/1602.05411} {arXiv:1602.05411
  [astro-ph.HE]} \BibitemShut {NoStop}%
\bibitem [{GWE()}]{GWEMAlerts}%
  \BibitemOpen
  \href@noop {} {\enquote {\bibinfo {title} {Identification and follow up of
  electromagnetic counterparts of gravitational wave candidate events},}\
  }\bibinfo {howpublished}
  {\url{http://www.ligo.org/scientists/GWEMalerts.php}},\ \bibinfo {note}
  {accessed: 2016-01-15}\BibitemShut {NoStop}%
\bibitem [{\citenamefont {Abbott}\ \emph
  {et~al.}(2016{\natexlab{d}})\citenamefont {Abbott} \emph
  {et~al.}}]{abbott2016detchar}%
  \BibitemOpen
  \bibfield  {author} {\bibinfo {author} {\bibfnamefont {B.~P.}\ \bibnamefont
  {Abbott}} \emph {et~al.} (\bibinfo {collaboration} {LIGO Scientific,
  Virgo}),\ }\href@noop {} {\  (\bibinfo {year} {2016}{\natexlab{d}})},\
  \Eprint {http://arxiv.org/abs/1602.03844} {arXiv:1602.03844 [gr-qc]}
  \BibitemShut {NoStop}%
\bibitem [{gst({\natexlab{a}})}]{gstlal}%
  \BibitemOpen
  \href@noop {} {\enquote {\bibinfo {title} {Gstlal},}\ }\bibinfo
  {howpublished}
  {\url{https://www.lsc-group.phys.uwm.edu/daswg/projects/gstlal.html}}
  ({\natexlab{a}}),\ \bibinfo {note} {accessed: 2015-07-01}\BibitemShut
  {NoStop}%
\bibitem [{gst({\natexlab{b}})}]{gstreamer}%
  \BibitemOpen
  \href@noop {} {\enquote {\bibinfo {title} {Gstreamer},}\ }\bibinfo
  {howpublished} {\url{https://gstreamer.freedesktop.org}}
  ({\natexlab{b}})\BibitemShut {NoStop}%
\bibitem [{lal()}]{lalsuite}%
  \BibitemOpen
  \href@noop {} {\enquote {\bibinfo {title} {Lalsuite},}\ }\bibinfo
  {howpublished}
  {\url{https://www.lsc-group.phys.uwm.edu/daswg/projects/lalsuite.html}},\
  \bibinfo {note} {accessed: 2015-07-01}\BibitemShut {NoStop}%
\bibitem [{\citenamefont {Abbott}\ \emph
  {et~al.}(2016{\natexlab{e}})\citenamefont {Abbott}, \citenamefont {Abbott},
  \citenamefont {Abbott}, \citenamefont {Abernathy}, \citenamefont {Acernese},
  \citenamefont {Ackley}, \citenamefont {Adams}, \citenamefont {Adams},
  \citenamefont {Addesso}, \citenamefont {Adhikari} \emph
  {et~al.}}]{abbott2016gw150914}%
  \BibitemOpen
  \bibfield  {author} {\bibinfo {author} {\bibfnamefont {B.}~\bibnamefont
  {Abbott}}, \bibinfo {author} {\bibfnamefont {R.}~\bibnamefont {Abbott}},
  \bibinfo {author} {\bibfnamefont {T.}~\bibnamefont {Abbott}}, \bibinfo
  {author} {\bibfnamefont {M.}~\bibnamefont {Abernathy}}, \bibinfo {author}
  {\bibfnamefont {F.}~\bibnamefont {Acernese}}, \bibinfo {author}
  {\bibfnamefont {K.}~\bibnamefont {Ackley}}, \bibinfo {author} {\bibfnamefont
  {C.}~\bibnamefont {Adams}}, \bibinfo {author} {\bibfnamefont
  {T.}~\bibnamefont {Adams}}, \bibinfo {author} {\bibfnamefont
  {P.}~\bibnamefont {Addesso}}, \bibinfo {author} {\bibfnamefont
  {R.}~\bibnamefont {Adhikari}},  \emph {et~al.},\ }\href@noop {} {\bibfield
  {journal} {\bibinfo  {journal} {Physical Review D}\ }\textbf {\bibinfo
  {volume} {93}},\ \bibinfo {pages} {122003} (\bibinfo {year}
  {2016}{\natexlab{e}})}\BibitemShut {NoStop}%
\bibitem [{\citenamefont {Abadie}\ \emph
  {et~al.}(2012{\natexlab{a}})\citenamefont {Abadie}, \citenamefont {Abbott},
  \citenamefont {Abbott}, \citenamefont {Abbott}, \citenamefont {Abernathy},
  \citenamefont {Accadia}, \citenamefont {Acernese}, \citenamefont {Adams},
  \citenamefont {Adhikari}, \citenamefont {Affeldt} \emph
  {et~al.}}]{abadie2012first}%
  \BibitemOpen
  \bibfield  {author} {\bibinfo {author} {\bibfnamefont {J.}~\bibnamefont
  {Abadie}}, \bibinfo {author} {\bibfnamefont {B.}~\bibnamefont {Abbott}},
  \bibinfo {author} {\bibfnamefont {R.}~\bibnamefont {Abbott}}, \bibinfo
  {author} {\bibfnamefont {T.}~\bibnamefont {Abbott}}, \bibinfo {author}
  {\bibfnamefont {M.}~\bibnamefont {Abernathy}}, \bibinfo {author}
  {\bibfnamefont {T.}~\bibnamefont {Accadia}}, \bibinfo {author} {\bibfnamefont
  {F.}~\bibnamefont {Acernese}}, \bibinfo {author} {\bibfnamefont
  {C.}~\bibnamefont {Adams}}, \bibinfo {author} {\bibfnamefont
  {R.}~\bibnamefont {Adhikari}}, \bibinfo {author} {\bibfnamefont
  {C.}~\bibnamefont {Affeldt}},  \emph {et~al.},\ }\href@noop {} {\bibfield
  {journal} {\bibinfo  {journal} {Astronomy \& Astrophysics}\ }\textbf
  {\bibinfo {volume} {541}},\ \bibinfo {pages} {A155} (\bibinfo {year}
  {2012}{\natexlab{a}})}\BibitemShut {NoStop}%
\bibitem [{\citenamefont {Klimenko}\ \emph {et~al.}(2008)\citenamefont
  {Klimenko} \emph {et~al.}}]{klimenko2008cwb}%
  \BibitemOpen
  \bibfield  {author} {\bibinfo {author} {\bibfnamefont {S.}~\bibnamefont
  {Klimenko}} \emph {et~al.},\ }\href@noop {} {\bibfield  {journal} {\bibinfo
  {journal} {Classical and Quantum Gravity}\ }\textbf {\bibinfo {volume}
  {25}},\ \bibinfo {pages} {114029} (\bibinfo {year} {2008})}\BibitemShut
  {NoStop}%
\bibitem [{\citenamefont {{Lynch}}\ \emph {et~al.}(2015)\citenamefont
  {{Lynch}}, \citenamefont {{Vitale}}, \citenamefont {{Essick}}, \citenamefont
  {{Katsavounidis}},\ and\ \citenamefont {{Robinet}}}]{lynch2015gwb}%
  \BibitemOpen
  \bibfield  {author} {\bibinfo {author} {\bibfnamefont {R.}~\bibnamefont
  {{Lynch}}}, \bibinfo {author} {\bibfnamefont {S.}~\bibnamefont {{Vitale}}},
  \bibinfo {author} {\bibfnamefont {R.}~\bibnamefont {{Essick}}}, \bibinfo
  {author} {\bibfnamefont {E.}~\bibnamefont {{Katsavounidis}}}, \ and\ \bibinfo
  {author} {\bibfnamefont {F.}~\bibnamefont {{Robinet}}},\ }\href@noop {} {\
  (\bibinfo {year} {2015})},\ \Eprint {http://arxiv.org/abs/1511.05955}
  {arXiv:1511.05955 [gr-qc]} \BibitemShut {NoStop}%
\bibitem [{\citenamefont {Abbott}\ \emph
  {et~al.}(2016{\natexlab{f}})\citenamefont {Abbott} \emph
  {et~al.}}]{TheLIGOScientific:2016uux}%
  \BibitemOpen
  \bibfield  {author} {\bibinfo {author} {\bibfnamefont {B.~P.}\ \bibnamefont
  {Abbott}} \emph {et~al.} (\bibinfo {collaboration} {Virgo, LIGO
  Scientific}),\ }\href@noop {} {\  (\bibinfo {year} {2016}{\natexlab{f}})},\
  \Eprint {http://arxiv.org/abs/1602.03843} {arXiv:1602.03843 [gr-qc]}
  \BibitemShut {NoStop}%
\bibitem [{\citenamefont {Buskulic}\ \emph {et~al.}(2010)\citenamefont
  {Buskulic}, \citenamefont {Collaboration}, \citenamefont {Collaboration}
  \emph {et~al.}}]{buskulic2010very}%
  \BibitemOpen
  \bibfield  {author} {\bibinfo {author} {\bibfnamefont {D.}~\bibnamefont
  {Buskulic}}, \bibinfo {author} {\bibfnamefont {L.~S.}\ \bibnamefont
  {Collaboration}}, \bibinfo {author} {\bibfnamefont {V.}~\bibnamefont
  {Collaboration}},  \emph {et~al.},\ }\href@noop {} {\bibfield  {journal}
  {\bibinfo  {journal} {Classical and Quantum Gravity}\ }\textbf {\bibinfo
  {volume} {27}},\ \bibinfo {pages} {194013} (\bibinfo {year}
  {2010})}\BibitemShut {NoStop}%
\bibitem [{\citenamefont {Babak}\ \emph {et~al.}(2013)\citenamefont {Babak},
  \citenamefont {Biswas}, \citenamefont {Brady}, \citenamefont {Brown},
  \citenamefont {Cannon}, \citenamefont {Capano}, \citenamefont {Clayton},
  \citenamefont {Cokelaer}, \citenamefont {Creighton}, \citenamefont {Dent}
  \emph {et~al.}}]{babak2013searching}%
  \BibitemOpen
  \bibfield  {author} {\bibinfo {author} {\bibfnamefont {S.}~\bibnamefont
  {Babak}}, \bibinfo {author} {\bibfnamefont {R.}~\bibnamefont {Biswas}},
  \bibinfo {author} {\bibfnamefont {P.}~\bibnamefont {Brady}}, \bibinfo
  {author} {\bibfnamefont {D.~A.}\ \bibnamefont {Brown}}, \bibinfo {author}
  {\bibfnamefont {K.}~\bibnamefont {Cannon}}, \bibinfo {author} {\bibfnamefont
  {C.~D.}\ \bibnamefont {Capano}}, \bibinfo {author} {\bibfnamefont {J.~H.}\
  \bibnamefont {Clayton}}, \bibinfo {author} {\bibfnamefont {T.}~\bibnamefont
  {Cokelaer}}, \bibinfo {author} {\bibfnamefont {J.~D.}\ \bibnamefont
  {Creighton}}, \bibinfo {author} {\bibfnamefont {T.}~\bibnamefont {Dent}},
  \emph {et~al.},\ }\href@noop {} {\bibfield  {journal} {\bibinfo  {journal}
  {Physical Review D}\ }\textbf {\bibinfo {volume} {87}},\ \bibinfo {pages}
  {024033} (\bibinfo {year} {2013})}\BibitemShut {NoStop}%
\bibitem [{\citenamefont {Cannon}\ \emph
  {et~al.}(2012{\natexlab{a}})\citenamefont {Cannon}, \citenamefont {Cariou},
  \citenamefont {Chapman}, \citenamefont {Crispin-Ortuzar}, \citenamefont
  {Fotopoulos}, \citenamefont {Frei}, \citenamefont {Hanna}, \citenamefont
  {Kara}, \citenamefont {Keppel}, \citenamefont {Liao} \emph
  {et~al.}}]{cannon2012toward}%
  \BibitemOpen
  \bibfield  {author} {\bibinfo {author} {\bibfnamefont {K.}~\bibnamefont
  {Cannon}}, \bibinfo {author} {\bibfnamefont {R.}~\bibnamefont {Cariou}},
  \bibinfo {author} {\bibfnamefont {A.}~\bibnamefont {Chapman}}, \bibinfo
  {author} {\bibfnamefont {M.}~\bibnamefont {Crispin-Ortuzar}}, \bibinfo
  {author} {\bibfnamefont {N.}~\bibnamefont {Fotopoulos}}, \bibinfo {author}
  {\bibfnamefont {M.}~\bibnamefont {Frei}}, \bibinfo {author} {\bibfnamefont
  {C.}~\bibnamefont {Hanna}}, \bibinfo {author} {\bibfnamefont
  {E.}~\bibnamefont {Kara}}, \bibinfo {author} {\bibfnamefont {D.}~\bibnamefont
  {Keppel}}, \bibinfo {author} {\bibfnamefont {L.}~\bibnamefont {Liao}},  \emph
  {et~al.},\ }\href@noop {} {\bibfield  {journal} {\bibinfo  {journal} {The
  Astrophysical Journal}\ }\textbf {\bibinfo {volume} {748}},\ \bibinfo {pages}
  {136} (\bibinfo {year} {2012}{\natexlab{a}})}\BibitemShut {NoStop}%
\bibitem [{\citenamefont {Allen}\ \emph {et~al.}(2012)\citenamefont {Allen},
  \citenamefont {Anderson}, \citenamefont {Brady}, \citenamefont {Brown},\ and\
  \citenamefont {Creighton}}]{allen2012findchirp}%
  \BibitemOpen
  \bibfield  {author} {\bibinfo {author} {\bibfnamefont {B.}~\bibnamefont
  {Allen}}, \bibinfo {author} {\bibfnamefont {W.~G.}\ \bibnamefont {Anderson}},
  \bibinfo {author} {\bibfnamefont {P.~R.}\ \bibnamefont {Brady}}, \bibinfo
  {author} {\bibfnamefont {D.~A.}\ \bibnamefont {Brown}}, \ and\ \bibinfo
  {author} {\bibfnamefont {J.~D.}\ \bibnamefont {Creighton}},\ }\href@noop {}
  {\bibfield  {journal} {\bibinfo  {journal} {Physical Review D}\ }\textbf
  {\bibinfo {volume} {85}},\ \bibinfo {pages} {122006} (\bibinfo {year}
  {2012})}\BibitemShut {NoStop}%
\bibitem [{\citenamefont {Allen}(2005)}]{allen2005chi}%
  \BibitemOpen
  \bibfield  {author} {\bibinfo {author} {\bibfnamefont {B.}~\bibnamefont
  {Allen}},\ }\href@noop {} {\bibfield  {journal} {\bibinfo  {journal}
  {Physical Review D}\ }\textbf {\bibinfo {volume} {71}},\ \bibinfo {pages}
  {062001} (\bibinfo {year} {2005})}\BibitemShut {NoStop}%
\bibitem [{\citenamefont {Cannon}\ \emph {et~al.}(2015)\citenamefont {Cannon},
  \citenamefont {Hanna},\ and\ \citenamefont {Peoples}}]{cannon2015likelihood}%
  \BibitemOpen
  \bibfield  {author} {\bibinfo {author} {\bibfnamefont {K.}~\bibnamefont
  {Cannon}}, \bibinfo {author} {\bibfnamefont {C.}~\bibnamefont {Hanna}}, \
  and\ \bibinfo {author} {\bibfnamefont {J.}~\bibnamefont {Peoples}},\
  }\href@noop {} {\bibfield  {journal} {\bibinfo  {journal} {arXiv preprint
  arXiv:1504.04632}\ } (\bibinfo {year} {2015})}\BibitemShut {NoStop}%
\bibitem [{\citenamefont {Cannon}\ \emph {et~al.}(2013)\citenamefont {Cannon},
  \citenamefont {Hanna},\ and\ \citenamefont {Keppel}}]{cannon2013method}%
  \BibitemOpen
  \bibfield  {author} {\bibinfo {author} {\bibfnamefont {K.}~\bibnamefont
  {Cannon}}, \bibinfo {author} {\bibfnamefont {C.}~\bibnamefont {Hanna}}, \
  and\ \bibinfo {author} {\bibfnamefont {D.}~\bibnamefont {Keppel}},\
  }\href@noop {} {\bibfield  {journal} {\bibinfo  {journal} {Physical Review
  D}\ }\textbf {\bibinfo {volume} {88}},\ \bibinfo {pages} {024025} (\bibinfo
  {year} {2013})}\BibitemShut {NoStop}%
\bibitem [{gra()}]{gracedb}%
  \BibitemOpen
  \href@noop {} {\enquote {\bibinfo {title} {Gravitational wave candidate event
  database},}\ }\bibinfo {howpublished}
  {\url{https://www.lsc-group.phys.uwm.edu/daswg/projects/gracedb.html}},\
  \bibinfo {note} {accessed: 2016-01-15}\BibitemShut {NoStop}%
\bibitem [{\citenamefont {Anderson}\ \emph {et~al.}(2009)\citenamefont
  {Anderson} \emph {et~al.}}]{FrameFormat}%
  \BibitemOpen
  \bibfield  {author} {\bibinfo {author} {\bibfnamefont {S.}~\bibnamefont
  {Anderson}} \emph {et~al.},\ }\href {{https://dcc.ligo.org/T970130/public}}
  {\enquote {\bibinfo {title} {{Specification of a Common Data Frame Format for
  Interferometric Gravitational Wave Detectors}},}\ } (\bibinfo {year}
  {2009}),\ \bibinfo {note} {{LIGO-DCC-T970130}}\BibitemShut {NoStop}%
\bibitem [{\citenamefont {Droz}\ \emph {et~al.}(1999)\citenamefont {Droz},
  \citenamefont {Knapp}, \citenamefont {Poisson},\ and\ \citenamefont
  {Owen}}]{droz1999gravitational}%
  \BibitemOpen
  \bibfield  {author} {\bibinfo {author} {\bibfnamefont {S.}~\bibnamefont
  {Droz}}, \bibinfo {author} {\bibfnamefont {D.~J.}\ \bibnamefont {Knapp}},
  \bibinfo {author} {\bibfnamefont {E.}~\bibnamefont {Poisson}}, \ and\
  \bibinfo {author} {\bibfnamefont {B.~J.}\ \bibnamefont {Owen}},\ }\href@noop
  {} {\bibfield  {journal} {\bibinfo  {journal} {Physical Review D}\ }\textbf
  {\bibinfo {volume} {59}},\ \bibinfo {pages} {124016} (\bibinfo {year}
  {1999})}\BibitemShut {NoStop}%
\bibitem [{\citenamefont {Abadie}\ \emph
  {et~al.}(2012{\natexlab{b}})\citenamefont {Abadie}, \citenamefont {Abbott},
  \citenamefont {Abbott}, \citenamefont {Abbott}, \citenamefont {Abernathy},
  \citenamefont {Accadia}, \citenamefont {Acernese}, \citenamefont {Adams},
  \citenamefont {Adhikari}, \citenamefont {Affeldt} \emph
  {et~al.}}]{abadie2012search_a}%
  \BibitemOpen
  \bibfield  {author} {\bibinfo {author} {\bibfnamefont {J.}~\bibnamefont
  {Abadie}}, \bibinfo {author} {\bibfnamefont {B.}~\bibnamefont {Abbott}},
  \bibinfo {author} {\bibfnamefont {R.}~\bibnamefont {Abbott}}, \bibinfo
  {author} {\bibfnamefont {T.}~\bibnamefont {Abbott}}, \bibinfo {author}
  {\bibfnamefont {M.}~\bibnamefont {Abernathy}}, \bibinfo {author}
  {\bibfnamefont {T.}~\bibnamefont {Accadia}}, \bibinfo {author} {\bibfnamefont
  {F.}~\bibnamefont {Acernese}}, \bibinfo {author} {\bibfnamefont
  {C.}~\bibnamefont {Adams}}, \bibinfo {author} {\bibfnamefont
  {R.}~\bibnamefont {Adhikari}}, \bibinfo {author} {\bibfnamefont
  {C.}~\bibnamefont {Affeldt}},  \emph {et~al.},\ }\href@noop {} {\bibfield
  {journal} {\bibinfo  {journal} {Physical Review D}\ }\textbf {\bibinfo
  {volume} {85}},\ \bibinfo {pages} {082002} (\bibinfo {year}
  {2012}{\natexlab{b}})}\BibitemShut {NoStop}%
\bibitem [{\citenamefont {Abadie}\ \emph
  {et~al.}(2012{\natexlab{c}})\citenamefont {Abadie}, \citenamefont {Abbott},
  \citenamefont {Abbott}, \citenamefont {Abbott}, \citenamefont {Abernathy},
  \citenamefont {Accadia}, \citenamefont {Acernese}, \citenamefont {Adams},
  \citenamefont {Adhikari}, \citenamefont {Affeldt} \emph
  {et~al.}}]{abadie2012search_b}%
  \BibitemOpen
  \bibfield  {author} {\bibinfo {author} {\bibfnamefont {J.}~\bibnamefont
  {Abadie}}, \bibinfo {author} {\bibfnamefont {B.}~\bibnamefont {Abbott}},
  \bibinfo {author} {\bibfnamefont {R.}~\bibnamefont {Abbott}}, \bibinfo
  {author} {\bibfnamefont {T.}~\bibnamefont {Abbott}}, \bibinfo {author}
  {\bibfnamefont {M.}~\bibnamefont {Abernathy}}, \bibinfo {author}
  {\bibfnamefont {T.}~\bibnamefont {Accadia}}, \bibinfo {author} {\bibfnamefont
  {F.}~\bibnamefont {Acernese}}, \bibinfo {author} {\bibfnamefont
  {C.}~\bibnamefont {Adams}}, \bibinfo {author} {\bibfnamefont
  {R.}~\bibnamefont {Adhikari}}, \bibinfo {author} {\bibfnamefont
  {C.}~\bibnamefont {Affeldt}},  \emph {et~al.},\ }\href@noop {} {\bibfield
  {journal} {\bibinfo  {journal} {The Astrophysical Journal}\ }\textbf
  {\bibinfo {volume} {760}},\ \bibinfo {pages} {12} (\bibinfo {year}
  {2012}{\natexlab{c}})}\BibitemShut {NoStop}%
\bibitem [{\citenamefont {Slutsky}\ \emph {et~al.}(2010)\citenamefont
  {Slutsky}, \citenamefont {Blackburn}, \citenamefont {Brown}, \citenamefont
  {Cadonati}, \citenamefont {Cain}, \citenamefont {Cavaglia}, \citenamefont
  {Chatterji}, \citenamefont {Christensen}, \citenamefont {Coughlin},
  \citenamefont {Desai} \emph {et~al.}}]{slutsky2010methods}%
  \BibitemOpen
  \bibfield  {author} {\bibinfo {author} {\bibfnamefont {J.}~\bibnamefont
  {Slutsky}}, \bibinfo {author} {\bibfnamefont {L.}~\bibnamefont {Blackburn}},
  \bibinfo {author} {\bibfnamefont {D.}~\bibnamefont {Brown}}, \bibinfo
  {author} {\bibfnamefont {L.}~\bibnamefont {Cadonati}}, \bibinfo {author}
  {\bibfnamefont {J.}~\bibnamefont {Cain}}, \bibinfo {author} {\bibfnamefont
  {M.}~\bibnamefont {Cavaglia}}, \bibinfo {author} {\bibfnamefont
  {S.}~\bibnamefont {Chatterji}}, \bibinfo {author} {\bibfnamefont
  {N.}~\bibnamefont {Christensen}}, \bibinfo {author} {\bibfnamefont
  {M.}~\bibnamefont {Coughlin}}, \bibinfo {author} {\bibfnamefont
  {S.}~\bibnamefont {Desai}},  \emph {et~al.},\ }\href@noop {} {\bibfield
  {journal} {\bibinfo  {journal} {Classical and Quantum Gravity}\ }\textbf
  {\bibinfo {volume} {27}},\ \bibinfo {pages} {165023} (\bibinfo {year}
  {2010})}\BibitemShut {NoStop}%
\bibitem [{\citenamefont {Christensen}\ \emph {et~al.}(2010)\citenamefont
  {Christensen}, \citenamefont {Collaboration}, \citenamefont {Collaboration}
  \emph {et~al.}}]{christensen2010ligo}%
  \BibitemOpen
  \bibfield  {author} {\bibinfo {author} {\bibfnamefont {N.}~\bibnamefont
  {Christensen}}, \bibinfo {author} {\bibfnamefont {L.~S.}\ \bibnamefont
  {Collaboration}}, \bibinfo {author} {\bibfnamefont {V.}~\bibnamefont
  {Collaboration}},  \emph {et~al.},\ }\href@noop {} {\bibfield  {journal}
  {\bibinfo  {journal} {Classical and Quantum Gravity}\ }\textbf {\bibinfo
  {volume} {27}},\ \bibinfo {pages} {194010} (\bibinfo {year}
  {2010})}\BibitemShut {NoStop}%
\bibitem [{\citenamefont {Owen}(1996)}]{owen1996search}%
  \BibitemOpen
  \bibfield  {author} {\bibinfo {author} {\bibfnamefont {B.~J.}\ \bibnamefont
  {Owen}},\ }\href@noop {} {\bibfield  {journal} {\bibinfo  {journal} {Physical
  Review D}\ }\textbf {\bibinfo {volume} {53}},\ \bibinfo {pages} {6749}
  (\bibinfo {year} {1996})}\BibitemShut {NoStop}%
\bibitem [{\citenamefont {Owen}\ and\ \citenamefont
  {Sathyaprakash}(1999)}]{owen1999matched}%
  \BibitemOpen
  \bibfield  {author} {\bibinfo {author} {\bibfnamefont {B.~J.}\ \bibnamefont
  {Owen}}\ and\ \bibinfo {author} {\bibfnamefont {B.}~\bibnamefont
  {Sathyaprakash}},\ }\href@noop {} {\bibfield  {journal} {\bibinfo  {journal}
  {Physical Review D}\ }\textbf {\bibinfo {volume} {60}},\ \bibinfo {pages}
  {022002} (\bibinfo {year} {1999})}\BibitemShut {NoStop}%
\bibitem [{\citenamefont {Apostolatos}(1995)}]{apostolatos1995search}%
  \BibitemOpen
  \bibfield  {author} {\bibinfo {author} {\bibfnamefont {T.~A.}\ \bibnamefont
  {Apostolatos}},\ }\href@noop {} {\bibfield  {journal} {\bibinfo  {journal}
  {Physical Review D}\ }\textbf {\bibinfo {volume} {52}},\ \bibinfo {pages}
  {605} (\bibinfo {year} {1995})}\BibitemShut {NoStop}%
\bibitem [{\citenamefont {Cokelaer}(2007)}]{cokelaer2007gravitational}%
  \BibitemOpen
  \bibfield  {author} {\bibinfo {author} {\bibfnamefont {T.}~\bibnamefont
  {Cokelaer}},\ }\href {\doibase 10.1103/PhysRevD.76.102004} {\bibfield
  {journal} {\bibinfo  {journal} {Phys. Rev.}\ }\textbf {\bibinfo {volume}
  {D76}},\ \bibinfo {pages} {102004} (\bibinfo {year} {2007})},\ \Eprint
  {http://arxiv.org/abs/0706.4437} {arXiv:0706.4437 [gr-qc]} \BibitemShut
  {NoStop}%
\bibitem [{\citenamefont {Abbott}\ \emph {et~al.}(2008)\citenamefont {Abbott},
  \citenamefont {Abbott}, \citenamefont {Adhikari}, \citenamefont {Agresti},
  \citenamefont {Ajith}, \citenamefont {Allen}, \citenamefont {Amin},
  \citenamefont {Anderson}, \citenamefont {Anderson}, \citenamefont {Arain}
  \emph {et~al.}}]{abbott2008search}%
  \BibitemOpen
  \bibfield  {author} {\bibinfo {author} {\bibfnamefont {B.}~\bibnamefont
  {Abbott}}, \bibinfo {author} {\bibfnamefont {R.}~\bibnamefont {Abbott}},
  \bibinfo {author} {\bibfnamefont {R.}~\bibnamefont {Adhikari}}, \bibinfo
  {author} {\bibfnamefont {J.}~\bibnamefont {Agresti}}, \bibinfo {author}
  {\bibfnamefont {P.}~\bibnamefont {Ajith}}, \bibinfo {author} {\bibfnamefont
  {B.}~\bibnamefont {Allen}}, \bibinfo {author} {\bibfnamefont
  {R.}~\bibnamefont {Amin}}, \bibinfo {author} {\bibfnamefont {S.}~\bibnamefont
  {Anderson}}, \bibinfo {author} {\bibfnamefont {W.}~\bibnamefont {Anderson}},
  \bibinfo {author} {\bibfnamefont {M.}~\bibnamefont {Arain}},  \emph
  {et~al.},\ }\href@noop {} {\bibfield  {journal} {\bibinfo  {journal}
  {Physical Review D}\ }\textbf {\bibinfo {volume} {78}},\ \bibinfo {pages}
  {042002} (\bibinfo {year} {2008})}\BibitemShut {NoStop}%
\bibitem [{\citenamefont {Harry}\ \emph {et~al.}(2014)\citenamefont {Harry},
  \citenamefont {Nitz}, \citenamefont {Brown}, \citenamefont {Lundgren},
  \citenamefont {Ochsner},\ and\ \citenamefont
  {Keppel}}]{harry2014investigating}%
  \BibitemOpen
  \bibfield  {author} {\bibinfo {author} {\bibfnamefont {I.~W.}\ \bibnamefont
  {Harry}}, \bibinfo {author} {\bibfnamefont {A.~H.}\ \bibnamefont {Nitz}},
  \bibinfo {author} {\bibfnamefont {D.~A.}\ \bibnamefont {Brown}}, \bibinfo
  {author} {\bibfnamefont {A.~P.}\ \bibnamefont {Lundgren}}, \bibinfo {author}
  {\bibfnamefont {E.}~\bibnamefont {Ochsner}}, \ and\ \bibinfo {author}
  {\bibfnamefont {D.}~\bibnamefont {Keppel}},\ }\href@noop {} {\bibfield
  {journal} {\bibinfo  {journal} {Physical Review D}\ }\textbf {\bibinfo
  {volume} {89}},\ \bibinfo {pages} {024010} (\bibinfo {year}
  {2014})}\BibitemShut {NoStop}%
\bibitem [{\citenamefont {Babak}(2008)}]{babak2008building}%
  \BibitemOpen
  \bibfield  {author} {\bibinfo {author} {\bibfnamefont {S.}~\bibnamefont
  {Babak}},\ }\href@noop {} {\bibfield  {journal} {\bibinfo  {journal}
  {Classical and Quantum Gravity}\ }\textbf {\bibinfo {volume} {25}},\ \bibinfo
  {pages} {195011} (\bibinfo {year} {2008})}\BibitemShut {NoStop}%
\bibitem [{\citenamefont {Harry}\ \emph {et~al.}(2009)\citenamefont {Harry},
  \citenamefont {Allen},\ and\ \citenamefont
  {Sathyaprakash}}]{harry2009stochastic}%
  \BibitemOpen
  \bibfield  {author} {\bibinfo {author} {\bibfnamefont {I.~W.}\ \bibnamefont
  {Harry}}, \bibinfo {author} {\bibfnamefont {B.}~\bibnamefont {Allen}}, \ and\
  \bibinfo {author} {\bibfnamefont {B.}~\bibnamefont {Sathyaprakash}},\
  }\href@noop {} {\bibfield  {journal} {\bibinfo  {journal} {Physical Review
  D}\ }\textbf {\bibinfo {volume} {80}},\ \bibinfo {pages} {104014} (\bibinfo
  {year} {2009})}\BibitemShut {NoStop}%
\bibitem [{\citenamefont {Manca}\ and\ \citenamefont
  {Vallisneri}(2010)}]{manca2010cover}%
  \BibitemOpen
  \bibfield  {author} {\bibinfo {author} {\bibfnamefont {G.~M.}\ \bibnamefont
  {Manca}}\ and\ \bibinfo {author} {\bibfnamefont {M.}~\bibnamefont
  {Vallisneri}},\ }\href@noop {} {\bibfield  {journal} {\bibinfo  {journal}
  {Physical Review D}\ }\textbf {\bibinfo {volume} {81}},\ \bibinfo {pages}
  {024004} (\bibinfo {year} {2010})}\BibitemShut {NoStop}%
\bibitem [{\citenamefont {Privitera}\ \emph {et~al.}(2014)\citenamefont
  {Privitera}, \citenamefont {Mohapatra}, \citenamefont {Ajith}, \citenamefont
  {Cannon}, \citenamefont {Fotopoulos}, \citenamefont {Frei}, \citenamefont
  {Hanna}, \citenamefont {Weinstein},\ and\ \citenamefont
  {Whelan}}]{privitera2014improving}%
  \BibitemOpen
  \bibfield  {author} {\bibinfo {author} {\bibfnamefont {S.}~\bibnamefont
  {Privitera}}, \bibinfo {author} {\bibfnamefont {S.~R.}\ \bibnamefont
  {Mohapatra}}, \bibinfo {author} {\bibfnamefont {P.}~\bibnamefont {Ajith}},
  \bibinfo {author} {\bibfnamefont {K.}~\bibnamefont {Cannon}}, \bibinfo
  {author} {\bibfnamefont {N.}~\bibnamefont {Fotopoulos}}, \bibinfo {author}
  {\bibfnamefont {M.~A.}\ \bibnamefont {Frei}}, \bibinfo {author}
  {\bibfnamefont {C.}~\bibnamefont {Hanna}}, \bibinfo {author} {\bibfnamefont
  {A.~J.}\ \bibnamefont {Weinstein}}, \ and\ \bibinfo {author} {\bibfnamefont
  {J.~T.}\ \bibnamefont {Whelan}},\ }\href@noop {} {\bibfield  {journal}
  {\bibinfo  {journal} {Physical Review D}\ }\textbf {\bibinfo {volume} {89}},\
  \bibinfo {pages} {024003} (\bibinfo {year} {2014})}\BibitemShut {NoStop}%
\bibitem [{\citenamefont {Cannon}\ \emph {et~al.}(2010)\citenamefont {Cannon},
  \citenamefont {Chapman}, \citenamefont {Hanna}, \citenamefont {Keppel},
  \citenamefont {Searle},\ and\ \citenamefont
  {Weinstein}}]{cannon2010singular}%
  \BibitemOpen
  \bibfield  {author} {\bibinfo {author} {\bibfnamefont {K.}~\bibnamefont
  {Cannon}}, \bibinfo {author} {\bibfnamefont {A.}~\bibnamefont {Chapman}},
  \bibinfo {author} {\bibfnamefont {C.}~\bibnamefont {Hanna}}, \bibinfo
  {author} {\bibfnamefont {D.}~\bibnamefont {Keppel}}, \bibinfo {author}
  {\bibfnamefont {A.~C.}\ \bibnamefont {Searle}}, \ and\ \bibinfo {author}
  {\bibfnamefont {A.~J.}\ \bibnamefont {Weinstein}},\ }\href@noop {} {\bibfield
   {journal} {\bibinfo  {journal} {Physical Review D}\ }\textbf {\bibinfo
  {volume} {82}},\ \bibinfo {pages} {044025} (\bibinfo {year}
  {2010})}\BibitemShut {NoStop}%
\bibitem [{\citenamefont {Cannon}\ \emph {et~al.}(2011)\citenamefont {Cannon},
  \citenamefont {Hanna},\ and\ \citenamefont {Keppel}}]{cannon2011efficiently}%
  \BibitemOpen
  \bibfield  {author} {\bibinfo {author} {\bibfnamefont {K.}~\bibnamefont
  {Cannon}}, \bibinfo {author} {\bibfnamefont {C.}~\bibnamefont {Hanna}}, \
  and\ \bibinfo {author} {\bibfnamefont {D.}~\bibnamefont {Keppel}},\
  }\href@noop {} {\bibfield  {journal} {\bibinfo  {journal} {Physical Review
  D}\ }\textbf {\bibinfo {volume} {84}},\ \bibinfo {pages} {084003} (\bibinfo
  {year} {2011})}\BibitemShut {NoStop}%
\bibitem [{\citenamefont {Cannon}\ \emph
  {et~al.}(2012{\natexlab{b}})\citenamefont {Cannon}, \citenamefont {Hanna},\
  and\ \citenamefont {Keppel}}]{cannon2012interpolating}%
  \BibitemOpen
  \bibfield  {author} {\bibinfo {author} {\bibfnamefont {K.}~\bibnamefont
  {Cannon}}, \bibinfo {author} {\bibfnamefont {C.}~\bibnamefont {Hanna}}, \
  and\ \bibinfo {author} {\bibfnamefont {D.}~\bibnamefont {Keppel}},\
  }\href@noop {} {\bibfield  {journal} {\bibinfo  {journal} {Physical Review
  D}\ }\textbf {\bibinfo {volume} {85}},\ \bibinfo {pages} {081504} (\bibinfo
  {year} {2012}{\natexlab{b}})}\BibitemShut {NoStop}%
\bibitem [{\citenamefont {Fairhurst}(2009)}]{fairhurst2009triangulation}%
  \BibitemOpen
  \bibfield  {author} {\bibinfo {author} {\bibfnamefont {S.}~\bibnamefont
  {Fairhurst}},\ }\href@noop {} {\bibfield  {journal} {\bibinfo  {journal} {New
  Journal of Physics}\ }\textbf {\bibinfo {volume} {11}},\ \bibinfo {pages}
  {123006} (\bibinfo {year} {2009})}\BibitemShut {NoStop}%
\bibitem [{\citenamefont {Singer}\ and\ \citenamefont
  {Price}(2016)}]{singer2016rapid}%
  \BibitemOpen
  \bibfield  {author} {\bibinfo {author} {\bibfnamefont {L.~P.}\ \bibnamefont
  {Singer}}\ and\ \bibinfo {author} {\bibfnamefont {L.~R.}\ \bibnamefont
  {Price}},\ }\href@noop {} {\bibfield  {journal} {\bibinfo  {journal}
  {Physical Review D}\ }\textbf {\bibinfo {volume} {93}},\ \bibinfo {pages}
  {024013} (\bibinfo {year} {2016})}\BibitemShut {NoStop}%
\bibitem [{\citenamefont {Robinson}\ \emph {et~al.}(2008)\citenamefont
  {Robinson}, \citenamefont {Sathyaprakash},\ and\ \citenamefont
  {Sengupta}}]{robinson2008geometric}%
  \BibitemOpen
  \bibfield  {author} {\bibinfo {author} {\bibfnamefont {C.}~\bibnamefont
  {Robinson}}, \bibinfo {author} {\bibfnamefont {B.}~\bibnamefont
  {Sathyaprakash}}, \ and\ \bibinfo {author} {\bibfnamefont {A.~S.}\
  \bibnamefont {Sengupta}},\ }\href@noop {} {\bibfield  {journal} {\bibinfo
  {journal} {Physical Review D}\ }\textbf {\bibinfo {volume} {78}},\ \bibinfo
  {pages} {062002} (\bibinfo {year} {2008})}\BibitemShut {NoStop}%
\bibitem [{\citenamefont {Capano}(2011)}]{capano2011searching}%
  \BibitemOpen
  \bibfield  {author} {\bibinfo {author} {\bibfnamefont {C.~D.}\ \bibnamefont
  {Capano}},\ }\emph {\bibinfo {title} {Searching for gravitational waves from
  compact binary coalescence using LIGO and virgo data}},\ \href@noop {} {Ph.D.
  thesis},\ \bibinfo  {school} {Syracuse University} (\bibinfo {year}
  {2011})\BibitemShut {NoStop}%
\bibitem [{lva()}]{lvalert}%
  \BibitemOpen
  \href@noop {} {}\bibinfo {howpublished}
  {\url{https://www.lsc-group.phys.uwm.edu/daswg/projects/lvalert.html}},\
  \bibinfo {note} {accessed: 2016-01-15}\BibitemShut {NoStop}%
\bibitem [{\citenamefont {Berry}\ \emph {et~al.}(2014)\citenamefont {Berry},
  \citenamefont {Mandel}, \citenamefont {Middleton}, \citenamefont {Singer},
  \citenamefont {Urban}, \citenamefont {Vecchio}, \citenamefont {Vitale},
  \citenamefont {Cannon}, \citenamefont {Farr}, \citenamefont {Farr} \emph
  {et~al.}}]{berry2014parameter}%
  \BibitemOpen
  \bibfield  {author} {\bibinfo {author} {\bibfnamefont {C.~P.}\ \bibnamefont
  {Berry}}, \bibinfo {author} {\bibfnamefont {I.}~\bibnamefont {Mandel}},
  \bibinfo {author} {\bibfnamefont {H.}~\bibnamefont {Middleton}}, \bibinfo
  {author} {\bibfnamefont {L.~P.}\ \bibnamefont {Singer}}, \bibinfo {author}
  {\bibfnamefont {A.~L.}\ \bibnamefont {Urban}}, \bibinfo {author}
  {\bibfnamefont {A.}~\bibnamefont {Vecchio}}, \bibinfo {author} {\bibfnamefont
  {S.}~\bibnamefont {Vitale}}, \bibinfo {author} {\bibfnamefont
  {K.}~\bibnamefont {Cannon}}, \bibinfo {author} {\bibfnamefont
  {B.}~\bibnamefont {Farr}}, \bibinfo {author} {\bibfnamefont {W.~M.}\
  \bibnamefont {Farr}},  \emph {et~al.},\ }\href@noop {} {\bibfield  {journal}
  {\bibinfo  {journal} {arXiv preprint arXiv:1411.6934}\ } (\bibinfo {year}
  {2014})}\BibitemShut {NoStop}%
\bibitem [{\citenamefont {Veitch}\ \emph {et~al.}(2015)\citenamefont {Veitch},
  \citenamefont {Raymond}, \citenamefont {Farr}, \citenamefont {Farr},
  \citenamefont {Graff}, \citenamefont {Vitale}, \citenamefont {Aylott},
  \citenamefont {Blackburn}, \citenamefont {Christensen}, \citenamefont
  {Coughlin} \emph {et~al.}}]{veitch2015parameter}%
  \BibitemOpen
  \bibfield  {author} {\bibinfo {author} {\bibfnamefont {J.}~\bibnamefont
  {Veitch}}, \bibinfo {author} {\bibfnamefont {V.}~\bibnamefont {Raymond}},
  \bibinfo {author} {\bibfnamefont {B.}~\bibnamefont {Farr}}, \bibinfo {author}
  {\bibfnamefont {W.}~\bibnamefont {Farr}}, \bibinfo {author} {\bibfnamefont
  {P.}~\bibnamefont {Graff}}, \bibinfo {author} {\bibfnamefont
  {S.}~\bibnamefont {Vitale}}, \bibinfo {author} {\bibfnamefont
  {B.}~\bibnamefont {Aylott}}, \bibinfo {author} {\bibfnamefont
  {K.}~\bibnamefont {Blackburn}}, \bibinfo {author} {\bibfnamefont
  {N.}~\bibnamefont {Christensen}}, \bibinfo {author} {\bibfnamefont
  {M.}~\bibnamefont {Coughlin}},  \emph {et~al.},\ }\href@noop {} {\bibfield
  {journal} {\bibinfo  {journal} {Physical Review D}\ }\textbf {\bibinfo
  {volume} {91}},\ \bibinfo {pages} {042003} (\bibinfo {year}
  {2015})}\BibitemShut {NoStop}%
\bibitem [{lig()}]{ligodv}%
  \BibitemOpen
  \href@noop {} {\enquote {\bibinfo {title} {Ligo data viewer},}\ }\bibinfo
  {howpublished}
  {\url{https://www.lsc-group.phys.uwm.edu/daswg/projects/ligodv.html}},\
  \bibinfo {note} {accessed: 2016-01-15}\BibitemShut {NoStop}%
\bibitem [{\citenamefont {Biswas}\ \emph {et~al.}(2013)\citenamefont {Biswas},
  \citenamefont {Blackburn}, \citenamefont {Cao}, \citenamefont {Essick},
  \citenamefont {Hodge}, \citenamefont {Katsavounidis}, \citenamefont {Kim},
  \citenamefont {Kim}, \citenamefont {Le~Bigot}, \citenamefont {Lee} \emph
  {et~al.}}]{biswas2013application}%
  \BibitemOpen
  \bibfield  {author} {\bibinfo {author} {\bibfnamefont {R.}~\bibnamefont
  {Biswas}}, \bibinfo {author} {\bibfnamefont {L.}~\bibnamefont {Blackburn}},
  \bibinfo {author} {\bibfnamefont {J.}~\bibnamefont {Cao}}, \bibinfo {author}
  {\bibfnamefont {R.}~\bibnamefont {Essick}}, \bibinfo {author} {\bibfnamefont
  {K.~A.}\ \bibnamefont {Hodge}}, \bibinfo {author} {\bibfnamefont
  {E.}~\bibnamefont {Katsavounidis}}, \bibinfo {author} {\bibfnamefont
  {K.}~\bibnamefont {Kim}}, \bibinfo {author} {\bibfnamefont {Y.-M.}\
  \bibnamefont {Kim}}, \bibinfo {author} {\bibfnamefont {E.-O.}\ \bibnamefont
  {Le~Bigot}}, \bibinfo {author} {\bibfnamefont {C.-H.}\ \bibnamefont {Lee}},
  \emph {et~al.},\ }\href@noop {} {\bibfield  {journal} {\bibinfo  {journal}
  {Physical Review D}\ }\textbf {\bibinfo {volume} {88}},\ \bibinfo {pages}
  {062003} (\bibinfo {year} {2013})}\BibitemShut {NoStop}%
\bibitem [{\citenamefont {Essick}\ \emph {et~al.}(2013)\citenamefont {Essick},
  \citenamefont {Blackburn},\ and\ \citenamefont
  {Katsavounidis}}]{essick2013optimizing}%
  \BibitemOpen
  \bibfield  {author} {\bibinfo {author} {\bibfnamefont {R.}~\bibnamefont
  {Essick}}, \bibinfo {author} {\bibfnamefont {L.}~\bibnamefont {Blackburn}}, \
  and\ \bibinfo {author} {\bibfnamefont {E.}~\bibnamefont {Katsavounidis}},\
  }\href@noop {} {\bibfield  {journal} {\bibinfo  {journal} {Classical and
  Quantum Gravity}\ }\textbf {\bibinfo {volume} {30}},\ \bibinfo {pages}
  {155010} (\bibinfo {year} {2013})}\BibitemShut {NoStop}%
\bibitem [{\citenamefont {Abadie}\ \emph
  {et~al.}(2010{\natexlab{b}})\citenamefont {Abadie}, \citenamefont {Abbott},
  \citenamefont {Abbott}, \citenamefont {Abernathy}, \citenamefont {Accadia},
  \citenamefont {Acernese}, \citenamefont {Adams}, \citenamefont {Adhikari},
  \citenamefont {Ajith}, \citenamefont {Allen} \emph
  {et~al.}}]{abadie2010sensitivity}%
  \BibitemOpen
  \bibfield  {author} {\bibinfo {author} {\bibfnamefont {J.}~\bibnamefont
  {Abadie}}, \bibinfo {author} {\bibfnamefont {B.}~\bibnamefont {Abbott}},
  \bibinfo {author} {\bibfnamefont {R.}~\bibnamefont {Abbott}}, \bibinfo
  {author} {\bibfnamefont {M.}~\bibnamefont {Abernathy}}, \bibinfo {author}
  {\bibfnamefont {T.}~\bibnamefont {Accadia}}, \bibinfo {author} {\bibfnamefont
  {F.}~\bibnamefont {Acernese}}, \bibinfo {author} {\bibfnamefont
  {C.}~\bibnamefont {Adams}}, \bibinfo {author} {\bibfnamefont
  {R.}~\bibnamefont {Adhikari}}, \bibinfo {author} {\bibfnamefont
  {P.}~\bibnamefont {Ajith}}, \bibinfo {author} {\bibfnamefont
  {B.}~\bibnamefont {Allen}},  \emph {et~al.},\ }\href@noop {} {\bibfield
  {journal} {\bibinfo  {journal} {arXiv preprint arXiv:1003.2481}\ } (\bibinfo
  {year} {2010}{\natexlab{b}})}\BibitemShut {NoStop}%
\bibitem [{\citenamefont {Meacher}\ \emph {et~al.}(2016)\citenamefont
  {Meacher}, \citenamefont {Cannon}, \citenamefont {Hanna}, \citenamefont
  {Regimbau},\ and\ \citenamefont {Sathyaprakash}}]{meacher2016second}%
  \BibitemOpen
  \bibfield  {author} {\bibinfo {author} {\bibfnamefont {D.}~\bibnamefont
  {Meacher}}, \bibinfo {author} {\bibfnamefont {K.}~\bibnamefont {Cannon}},
  \bibinfo {author} {\bibfnamefont {C.}~\bibnamefont {Hanna}}, \bibinfo
  {author} {\bibfnamefont {T.}~\bibnamefont {Regimbau}}, \ and\ \bibinfo
  {author} {\bibfnamefont {B.}~\bibnamefont {Sathyaprakash}},\ }\href@noop {}
  {\bibfield  {journal} {\bibinfo  {journal} {Physical Review D}\ }\textbf
  {\bibinfo {volume} {93}},\ \bibinfo {pages} {024018} (\bibinfo {year}
  {2016})}\BibitemShut {NoStop}%
\end{thebibliography}
